\documentclass[preprintnumbers, prd, showpacs, floatfix, onecolumn, preprintnumbers, letterpaper, superscriptaddress,nofootinbib]{revtex4}
\usepackage{eurosym}
\usepackage{amsfonts}
\usepackage{amsmath}
\usepackage{amssymb,epsf}
\usepackage{latexsym}
\usepackage{graphicx,epsfig}
\usepackage{amssymb}
\usepackage{float}
\usepackage{subfigure}
\usepackage[colorlinks=true,citecolor=blue,linkcolor=blue,urlcolor=black]{hyperref}
\usepackage{epstopdf}

\graphicspath{{Images/}}

\begin{document}

\title{Thermodynamics and reentrant phase transition\\
for logarithmic nonlinear charged black holes in massive gravity}
\author{S. Rajaee Chaloshtary}
\affiliation{Physics Department and Biruni Observatory, Shiraz University, Shiraz 71454,
Iran}
\author{M. Kord Zangeneh}
\email{mkzangeneh@scu.ac.ir}
\affiliation{Physics Department, Faculty of Science, Shahid Chamran University of Ahvaz,
Ahvaz 61357-43135, Iran}
\author{S. Hajkhalili}
\affiliation{Physics Department and Biruni Observatory, Shiraz University, Shiraz 71454,
Iran}
\author{A. Sheykhi}
\email{asheykhi@shirazu.ac.ir}
\affiliation{Physics Department and Biruni Observatory, Shiraz University, Shiraz 71454,
Iran}
\affiliation{Max-Planck-Institute for Gravitational Physics (Albert-Einstein-Institute),
14476 Potsdam, Germany}
\author{S. M. Zebarjad}
\email{zebarjad@shirazu.ac.ir}
\affiliation{Physics Department and Biruni Observatory, Shiraz University, Shiraz 71454,
Iran}

\begin{abstract}
We investigate a new class of $(n+1)$-dimensional topological black hole
solutions in the context of massive gravity and in the presence of
logarithmic nonlinear electrodynamics. Exploring higher dimensional
solutions in massive gravity coupled to nonlinear electrodynamics is
motivated by holographic hypothesis as well as string theory. We first
construct exact solutions of the field equations and then explore the
behavior of the metric functions for different values of the model
parameters. We observe that our black holes admit the multi-horizons caused
by a quantum effect called anti-evaporation. Next, by calculating the
conserved and thermodynamic quantities, we obtain a generalized Smarr
formula. We find that the first law of black holes thermodynamics is
satisfied on the black hole horizon. We study thermal stability of the
obtained solutions in both canonical and grand canonical ensembles. We
reveal that depending on the model parameters, our solutions exhibit a rich
variety of phase structures. Finally, we explore, for the first time without
extending thermodynamics phase space, the critical behavior and reentrant
phase transition for black hole solutions in massive gravity theory. We
realize that there is a zeroth order phase transition for a specified range
of charge value and the system experiences a large/small/large reentrant
phase transition due to the presence of nonlinear electrodynamics.
\end{abstract}

\maketitle

\section{INTRODUCTION}

The initial version of General Relativity (GR) presented by Einstein is a
theory which predicts the existence of a massless spin-$2$ particle called
graviton \cite{ref1}. One of the most natural generalizations of massless GR
theory is to admit the existence of the massive graviton and hence
constructing a massive theory of gravity. According to the particle physics,
Higgs mechanism bestows mass to carrier particles of the electroweak
interaction. Strong motivation for studying the massive gravity models comes
from recent developments in gravitational waves observatory. The recent
observation of the the gravitational wave event GW170817 and of its
electromagnetic counterpart GRB170817A, from a binary neutron star merger,
confirmed that the speed of gravitational waves deviates from the speed of
light by less than one part in $10^{15}$ \cite{Abbot1}. This implies that
one should consider an upper bound for the mass of graviton \cite{Abbot2}.
Thus the observation of the gravitational waves put severe constraints on
several theories of modified gravity. Any modified gravity model predicting
the speed of gravitational waves less than the speed of light must now be
seriously reconsidered \cite{Ales}. From holographic point of view, massive
gravity could make momentum dissipative in dual systems \cite{ref8}. Giving
mass to graviton state breaks the diffeomorphism invariance of stress-energy
tensor of dual theory holographically which in turn cause momentum
dissipation \cite{ref8}.

Historically, the first massive gravity theory was derived by Fierz and
Pauli by adding the interaction terms to the linearized level of GR \cite%
{ref3}. Their theory suffered a discontinuity in the predictions. For
instance, Van Dam, Veltman and Zakharov found a discontinuity in the
Newtonian potential in massless limit, recognized as VDVZ problem \cite{ref4}%
. Boulware and Deser suggested a nonlinear model \cite{ref5}, while
Vainshtein introduced the linearized theory as culprit for VDVZ problem \cite%
{ref16}. However, their theory was plagued with a ghost appeared as a 6th
degree of freedom. A four-dimensional covariant nonlinear theories of
massive gravity which are ghost-free in the decoupling limit to all orders
was established by de Rham-Gabadadze-Tolley (dRGT) \cite{ref6}. Their
theories resum explicitly all the nonlinear terms of an effective field
theory of massive gravity \cite{ref6}. Actually, the absence of higher
derivative terms in equations of motion does not allow the ghost to exist.
This theory has been explored from different point of view and known as the
only candidate for pathology-free nonlinear massive gravity so far \cite%
{ref13}. Although, finding exact solutions is not easy in this model due to
the complexity of equations, it has always been an interesting and well
motivated topic of research in theoretical physics \cite{ref11}.

Black hole solutions in the context of massive gravity have been explored
from different points of view. A Schwarzschild-de Sitter-like vacuum
solution of dRGT model has been found in \cite{ref7}. A nontrivial black
hole solution with a negative cosmological constant has been obtained in 
\cite{ref8}. It has been shown that the mass of the graviton in massive
theory is equivalent to the lattice in the holographic conductor model \cite%
{ref8} (see also \cite{ref9}). Charged black hole solutions of massive
gravity theory with negative cosmological constant have been constructed in 
\cite{ref10} and their thermodynamic properties and phase structure have
been studied in both canonical and grand canonical ensembles. Exact
three-dimensional asymptotically AdS-like solutions in massive gravity have
been explored in \cite{ref12}. The authors of Ref. \cite{ref17} found the
classical solutions for stars and other compact objects in massive gravity
following the Vainshtein mechanism. Neutron stars in the context of this
theory have been investigated in \cite{ref18}.

The importance of massive gravity does not restrict to the non-perturbative
aspects of gravity or black hole solutions. In the cosmological framework,
the massive gravity can be regarded as an alternative for the late time
cosmic acceleration \cite{cosmicAcc}. Thanks to Vainshtein mechanism,
massive gravity is equivalent to GR at small scales, although in large
distances (i.e. infrared regime) it amends gravity. Thus, the nature of the
cosmological constant may change since it is the most universal infrared
possible source \cite{ref14}. Furthermore, massive particles generate Yukawa
potential $\sim {e^{-m\,r}}/{r}$ which has an exponential suppression
kicking in at the length scale $\sim {1}/{m}$. If we imagine that $m\sim H$
where $H$ is Hubble scale, then the force of massive graviton would be
weakened at large scales and it may offer a nice solution for acceleration
problem \cite{ref15}. More researches in this regard can be found in \cite%
{ref19}.

When the linear Maxwell theory failed to explain the central singularity of
a point charge, the striking alternative was nonlinear electrodynamics
(NED). In $1930$, Born and Infeld (BI) made the first efforts to construct a
NED theory \cite{ref20}. Some years later, it turned out that their theory
plays a prominent role in D-brane physics. In the framework of particle
physics, NED was propounded by Heisenberg and Euler \cite{ref21}. Its
extension in GR was done by Plebanski \cite{ref22}. The theories of NED have
arisen a lot of attention after development in the string theory \cite{ref23}%
. For instance, loop calculations in open superstring theory lead to a low
energy effective action included BI type term \cite{ref26}. It is noteworthy
to mention that at the high energy levels, due to the interactions with
other physical fields, real electromagnetic field cannot obey linear law.
Basically, it was argued that NEDs are the simplified phenomenological
descriptions of the pointed out interactions. In addition, NED can explain
the Rindler acceleration as a nonlinear electromagnetic effect \cite{ref24}.
From quantum gravity viewpoint, NEDs are corrections to Maxwell field. So, a
large class of regular black holes (i.e. black holes without singularity)
have been studied in the literature. The pioneer effort in this direction is
attributed to Bardeen \cite{ref25}.

The Lagrangian of a general theory of NED is assumed to be a function of the
Maxwell invariant $\mathcal{F}=F_{\mu \nu }F^{\mu \nu }$ where $F_{\mu \nu }$
is electromagnetic tensor and not to contain any higher derivative terms of $%
\mathcal{F}$. In the weak field limit, these theories lead to linear Maxwell
one and so their expansions in terms of $\mathcal{F}$ have the form $%
\mathcal{L}(\mathcal{F})=\mathcal{F}+O(\mathcal{F}^{2})$. There has been
some efforts to construct BI type NED Lagrangian in literature. Two famous
ones are so-called Logarithmic (LNE) and Exponential (ENE) nonlinear
electrodynamics \cite{ref27,ref28}. LNE removes divergences in the electric
field (like BI model), whereas the ENE just makes it weaker than Maxwell
theory. In the present work, we are interested in considering LNE
Lagrangian. It is notable to mention that logarithmic term of the field
strength may come out as an exact one-loop correction to the vacuum
polarization. This fact has been proved by Euler and Heisenberg when they
studied electrons in a background setup by a uniform electromagnetic field 
\cite{ref21}.

LNE Lagrangian has been studied in various contexts. In Einstein-dilaton
theory, black hole solutions have been constructed and their thermodynamic
stability have been discussed \cite{ref30}. AdS-dilaton black holes in the
presence of LNE have been investigated in \cite{ref31}. Lifshitz-dilaton
black holes/branes coupled to LNE have been investigated in \cite{ref32}.
Magnetic rotating dilaton strings and charged rotating dilaton black strings
in Einstein gravity with LNE electrodynamics have been found in \cite{ref33}
and \cite{ref34}, respectively. In the context of Lovelock gravity, authors
of \cite{ref35} explored magnetic brane solutions in the presence of LNE.
Other studies on the black hole solutions in the presence of LNE can be
carried out in \cite{ref38}.

The layout of this paper is as follows. In the next section, we introduce
the Lagrangian of massive gravity in the presence of cosmological constant
and LNE and construct exact black hole solutions. We calculate conserved and
thermodynamic quantities of the solutions in section \ref{Therm}. Thermal
stability of the solutions is checked in \ref{Stability}. In \ref{Reentrant}%
, the critical behavior of ($3+1$)-dimensional black holes is examined. We
devoted the last section to summary and some closing remarks.

\section{Action and massive gravity solutions\label{lifsol}}

The ($n+1$)-dimensional action of Einstein massive gravity with a negative
cosmological constant and NLE is \cite{maldacena}

\begin{equation}
\mathcal{S}=\frac{1}{16\pi }\int d^{n+1}x\sqrt{-g}\left[ \mathcal{R}%
-2\Lambda -8\beta ^{2}\ln \left( 1+\frac{\mathcal{F}}{8\beta ^{2}}\right)
+m^{2}\sum_{i}^{4}c_{i}\mathcal{U}_{i}(g,\Gamma )\right] ,  \label{Action}
\end{equation}%
where $\mathcal{R}$ is the Ricci scalar, $\Lambda =-n(n-1)/2l^{2}$ with $l$
is the radius of AdS spacetime and $\mathcal{F}=F_{\mu \nu }F^{\mu \nu }$ is
the Maxwell invariant and $F_{\mu \nu }=2\partial _{\lbrack \mu }A_{\nu ]}$
is the electromagnetic field tensor, while $A_{\nu \text{ \ }}$ is the
electromagnetic vector potential. The constant $\beta $ is the parameter of
nonlinearity so that the logarithmic nonlinear electrodynamics recovers the
linear Maxwell theory when $\beta \rightarrow \infty $. $m^{2}$ is the
positive massive gravity parameter so that the translational invariance is
recovered as $m$ approachs to zero $(m\rightarrow 0).$

In action (\ref{Action}), $c_{i}$'s are constants and $\mathcal{U}_{i}$'s
are symmetric polynomials eigenvalues of $(n+1)\times (n+1)$ matrix $%
\mathcal{K}_{\nu }^{\mu }=\sqrt{g^{\mu \lambda }\Gamma _{\lambda \nu }}$\ so
that

\begin{eqnarray}
\mathcal{U}_{1} &=&\left[ \mathcal{K}\right] , \\
\mathcal{U}_{2} &=&\left[ \mathcal{K}\right] ^{2}-\left[ \mathcal{K}^{2}%
\right] , \\
\mathcal{U}_{3} &=&\left[ \mathcal{K}\right] ^{3}-3\left[ \mathcal{K}\right] %
\left[ \mathcal{K}^{2}\right] +2\left[ \mathcal{K}^{3}\right] , \\
\mathcal{U}_{4} &=&\left[ \mathcal{K}\right] ^{4}-6\left[ \mathcal{K}^{2}%
\right] \left[ \mathcal{K}\right] ^{2}+8\left[ \mathcal{K}^{3}\right] \left[ 
\mathcal{K}\right] +3\left[ \mathcal{K}^{2}\right] ^{2}-6\left[ \mathcal{K}%
^{4}\right] .
\end{eqnarray}%
Here, rectangular braket $\left[ \mathcal{K}\right] =\mathcal{K}_{\mu }^{\mu
}$ is the trace of the matrix square root which is defined as $(\sqrt{%
\mathcal{K}_{\nu }^{\mu }})\times (\sqrt{\mathcal{K}_{\lambda }^{v}})=%
\mathcal{K}_{\lambda }^{\mu }$. The dynamical metric is denoted by $g$ and $%
\Gamma $ is reference metric which is a $2$-rank symmetric tensor.

One can obtain the equations of motion by varying the action (\ref{Action})
with respect to the metric tensor $g_{\mu \nu }$ and gauge field $A_{\mu }$
as%
\begin{equation}
G_{\mu \nu }+\Lambda g_{\mu \nu }+\frac{2F_{\mu \lambda }F_{\nu }^{\text{ \ }%
\lambda }}{1+\frac{\mathcal{F}}{8\beta ^{2}}}+4\beta ^{2}\ln \left( 1+\frac{%
\mathcal{F}}{8\beta ^{2}}\right) g_{\mu \nu }+m^{2}\chi _{\mu \nu }=0,
\label{Feq1}
\end{equation}%
\begin{equation}
\nabla _{\mu }\left[ \left( 1+\frac{\mathcal{F}}{8\beta ^{2}}\right)
^{-1}F^{\mu \nu }\right] =0,  \label{Feq2}
\end{equation}%
where, $G_{\mu \nu }$ is the Einstein tensor and 
\begin{eqnarray}
\chi _{\mu \nu } &=&-\frac{c_{1}}{2}\left( \mathcal{U}_{1}g_{\mu \nu }-%
\mathcal{K}_{\mu \nu }\right) -\frac{c_{2}}{2}\left( \mathcal{U}_{2}g_{\mu
\nu }-2\mathcal{U}_{1}\mathcal{K}_{\mu \nu }+2\mathcal{K}_{\mu \nu
}^{2}\right) -\frac{c_{3}}{2}(\mathcal{U}_{3}g_{\mu \nu }-3\mathcal{U}_{2}%
\mathcal{K}_{\mu \nu } \\
&&+6\mathcal{U}_{1}\mathcal{K}_{\mu \nu }^{2}-6\mathcal{K}_{\mu \nu }^{3})-%
\frac{c_{4}}{2}(\mathcal{U}_{4}g_{\mu \nu }-4\mathcal{U}_{3}\mathcal{K}_{\mu
\nu }+12\mathcal{U}_{2}\mathcal{K}_{\mu \nu }^{2}-24\mathcal{U}_{1}\mathcal{K%
}_{\mu \nu }^{3}+24\mathcal{K}_{\mu \nu }^{4}).
\end{eqnarray}
In order to attain the static charged black hole solution, we consider the
line element of ($n+1$)-dimensional spacetime as 
\begin{equation}
ds^{2}=-f(r)dt^{2}+f^{-1}(r)dr^{2}+r^{2}h_{ij}(x)dx^{i}dx^{j},\qquad \qquad
\qquad (i,j=1,2,...n-1)  \label{metr}
\end{equation}%
where the function $f(r)$ should be determined and $h_{ij}(x)dx^{i}dx^{j}$
exhibits a hypersurface with constant curvature $(n-1)(n-2)k$ where $k$
explicate the topology of event horizon or the boundary of $t=constant$ - $%
r=constant$. It will be $k=0$ $(R^{n-1})$, $k=1$ $(S^{n-1})$ or $k=-1$ $%
(H^{n-1})$ . By taking $k=0,1,-1$, the black hole horizon can be flat
(zero), spherical (positive) or hyperbolic (negative) constant curvature
hypersurfaces with volume $V_{n-1}$. The reference metric can be considered
as \cite{refmetric} 
\begin{equation}
\Gamma _{\mu \nu }=\mathrm{diag}(0,0,c_{0}^{2}h_{ij}(x)),  \label{metref}
\end{equation}%
where $c_{0}$ is a positive constant. By employing metric (\ref{metr}) and (%
\ref{metref}) we can calculate $\mathcal{U}_{i}$ 's

\begin{eqnarray}
\mathcal{U}_{1} &=&\frac{(n-1)c_{0}}{r},  \notag  \label{Us} \\
\mathcal{U}_{2} &=&\frac{(n-1)(n-2)c_{0}^{2}}{r^{2}},  \notag \\
\mathcal{U}_{3} &=&\frac{(n-1)(n-2)(n-3)c_{0}^{3}}{r^{3}},  \notag \\
\mathcal{U}_{4} &=&\frac{(n-1)(n-2)(n-3)(n-3)c_{0}^{4}}{r^{4}},
\end{eqnarray}%
It is remarkable to note that $\mathcal{U}_{3}$ and $\mathcal{U}_{4}$ become
zero for $(3+1)$-dimensional spacetime and $\mathcal{U}_{4}$ vanishes for $%
(4+1)$-dimensional spacetime. By considering the metric (\ref{metr}), one
can integrate (\ref{Feq2}) to calculate the electromagnetic field

\begin{equation}
F_{tr}=-F_{rt}=\frac{2qr^{1-n}}{1+\Upsilon },  \label{Ftr}
\end{equation}%
with 
\begin{equation*}
\Upsilon =\sqrt{1+\left( \frac{q}{\beta r^{n-1}}\right) ^{2}}.
\end{equation*}%
In latter equations, $q$ is a constant which has relation with total charge
of black hole. Substituting the reference metric (\ref{metref}), (\ref{Us})
and electromagnetic field (\ref{Ftr}) into the field Eqs. (\ref{Feq1}), we
find

\begin{eqnarray}
f^{\prime }+\frac{(n-2)(f-k)}{r}+\frac{2\Lambda r}{n-1}-\frac{8\beta ^{2}r}{%
n-1}\left[ \ln \left( 1+\Upsilon \right) +1-\Upsilon \right]
-c_{0}m^{2}\left( c_{1}+\frac{(n-2)c_{0}c_{2}}{r}\right. &&  \notag \\
\left. +\frac{(n-2)(n-3)c_{0}^{2}c_{3}}{r^{2}}+\frac{%
(n-2)(n-3)(n-4)c_{0}^{3}c_{4}}{r^{3}}\right) &=&0,
\end{eqnarray}

\begin{eqnarray}
f^{\prime \prime }+\frac{2(n-2)}{r}f^{\prime }+\frac{(n-2)(n-3)(f+k)}{r^{2}}%
+2\Lambda -8\beta ^{2}\ln \left( \frac{1+\Upsilon }{2}\right) -\frac{\left(
n-2\right) c_{0}m^{2}}{r}\left( c_{1}+\frac{(n-3)c_{0}c_{2}}{r}\right. && 
\notag \\
\left. +\frac{(n-3)(n-4)c_{0}^{2}c_{3}}{r^{2}}+\frac{%
(n-3)(n-4)(n-5)c_{0}^{3}c_{4}}{r^{3}}\right) &=&0,
\end{eqnarray}%
where the prime denotes derivative with respect to $r$. Finally, one could
solve the above differential equations to obtain the metric function $f(r)$
as 
\begin{eqnarray}
f(r) &=&k-\frac{m_{0}}{r^{n-2}}-\frac{2r^{2}\Lambda }{n(n-1)}+\frac{8\beta
^{2}r^{2}}{n\left( n-1\right) }\left[ \ln \left( \frac{1+\Upsilon }{2}%
\right) +\frac{\left( 2n-1\right) \left( 1-\Upsilon \right) }{n}\right] 
\notag  \label{f(r)} \\
&&+\frac{8\left( n-1\right) q^{2}r^{4-2n}}{n^{2}\left( n-2\right) }\mathbf{F}%
\left( \frac{1}{2},\frac{-2+n}{2(-1+n)},\frac{4-3n}{2-2n},\frac{%
-q^{2}r^{2-2n}}{\beta ^{2}}\right)  \notag \\
&&+\frac{c_{0}m^{2}r}{n-1}\left( c_{1}+\frac{(n-1)c_{0}c_{2}}{r}+\frac{%
(n-1)(n-2)c_{0}^{2}c_{3}}{r^{2}}+\frac{(n-1)(n-2)(n-3)c_{0}^{3}c_{4}}{r^{3}}%
\right),
\end{eqnarray}%
where $\mathbf{F}$ is the hypergeometric function. Linear limit ($\beta
\rightarrow \infty $) of the metric function is 
\begin{eqnarray}
f(r) &=&k-\frac{m_{0}}{r^{n-2}}-\frac{2\Lambda r^{2}}{n(n-1)}+\frac{2q^{2}}{%
(n-1)(n-2)r^{2(n-2)}}-\frac{q^{4}r^{6-4n}}{4\left( n-1\right) \left(
3n-4\right) \beta ^{2}}  \notag \\
&&+\frac{c_{0}m^{2}r}{n-1}\left( c_{1}+\frac{(n-1)c_{0}c_{2}}{r}+\frac{%
(n-1)(n-2)c_{0}^{2}c_{3}}{r^{2}}+\frac{(n-1)(n-2)(n-3)c_{0}^{3}c_{4}}{r^{3}}%
\right) +O\left( \frac{1}{\beta ^{4}}\right)
\end{eqnarray}%
which is in accordance with the metric function of charged black hole in
Maxwell theory \cite{ref41} along with a correction term corresponding to
the nonlinear electrodynamics. 
\begin{figure}[t]
\centering
\subfigure[~$n=4$ and
    $q=1$]{\includegraphics[scale=0.3]{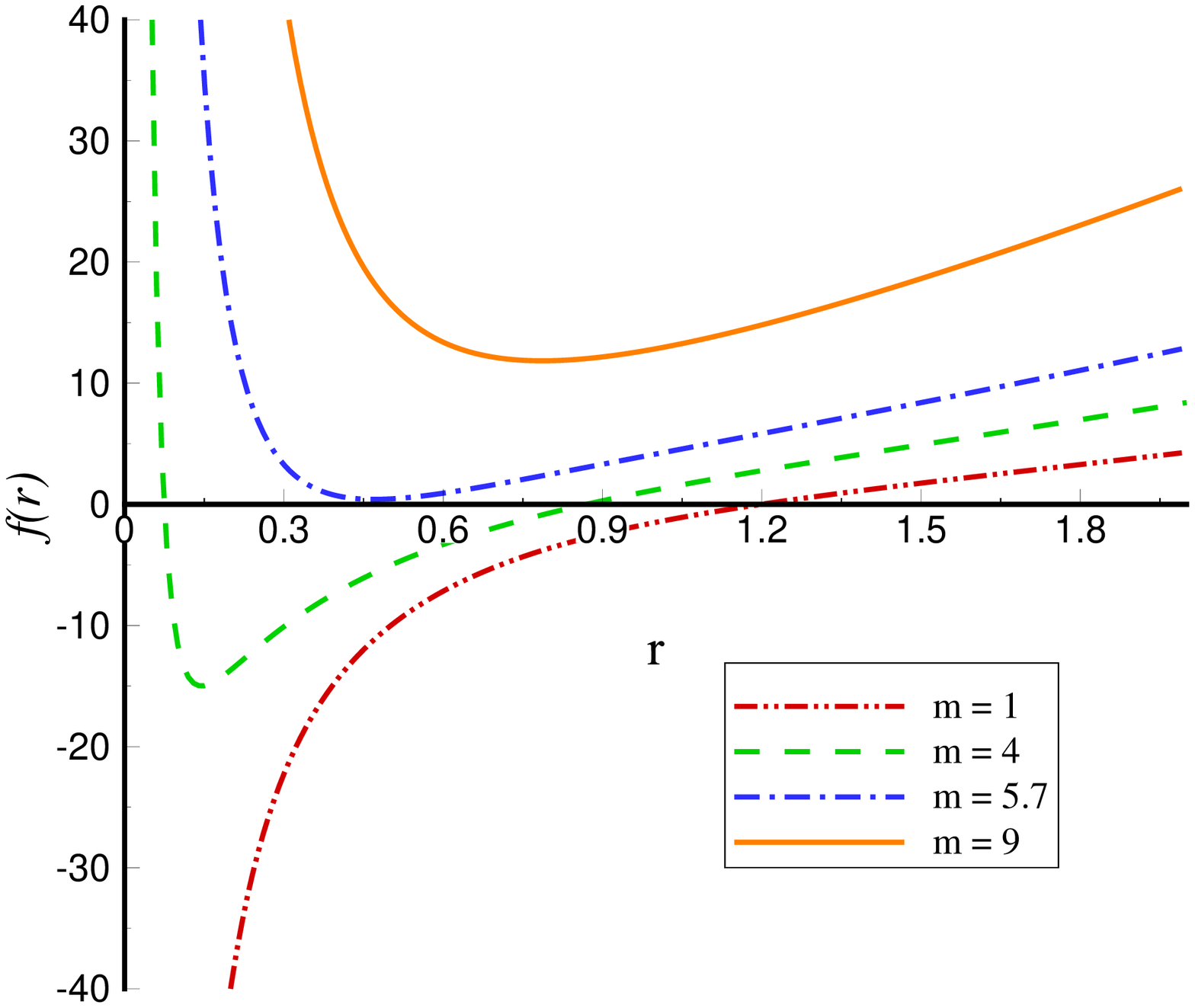}\label{fig1a}} \hspace*{0.5cm} 
\subfigure[~$n=3$ and
$m=4$]{\includegraphics[scale=0.3]{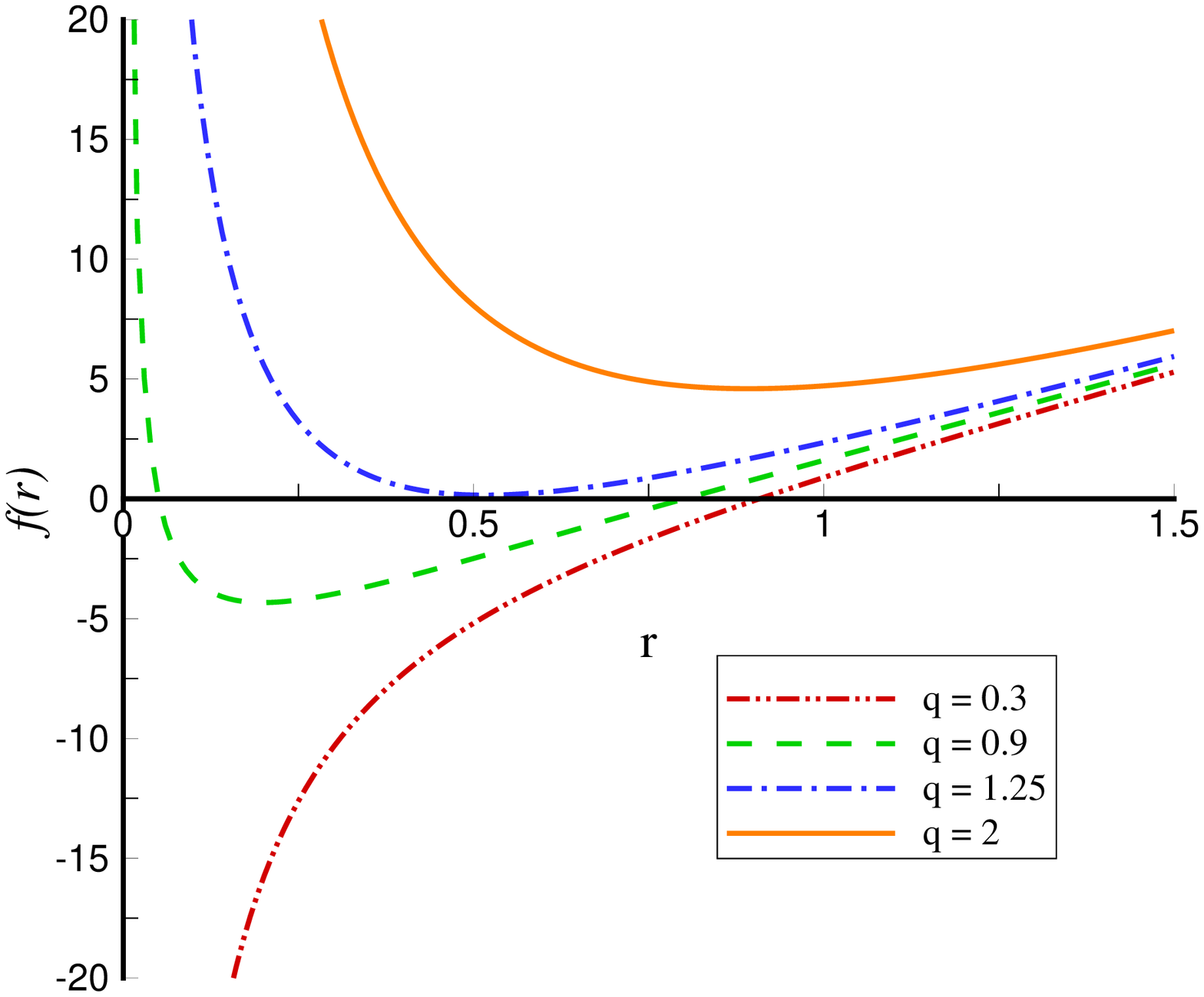}\label{fig1b}} 
\caption{The behavior of $f(r)$ versus $r$ for $\protect\beta =2$, $m_{0}=4$
, $c_{0}=0.5$, $c_{1}=1$, $c_{2}=-0.4$, $c_{3}=0.4$ and $c_{4}=0.5$}
\label{fig1}
\end{figure}
\begin{figure}[t]
\begin{minipage}[t]{0.4\textwidth}
        \includegraphics
        [width=\linewidth,keepaspectratio=true]{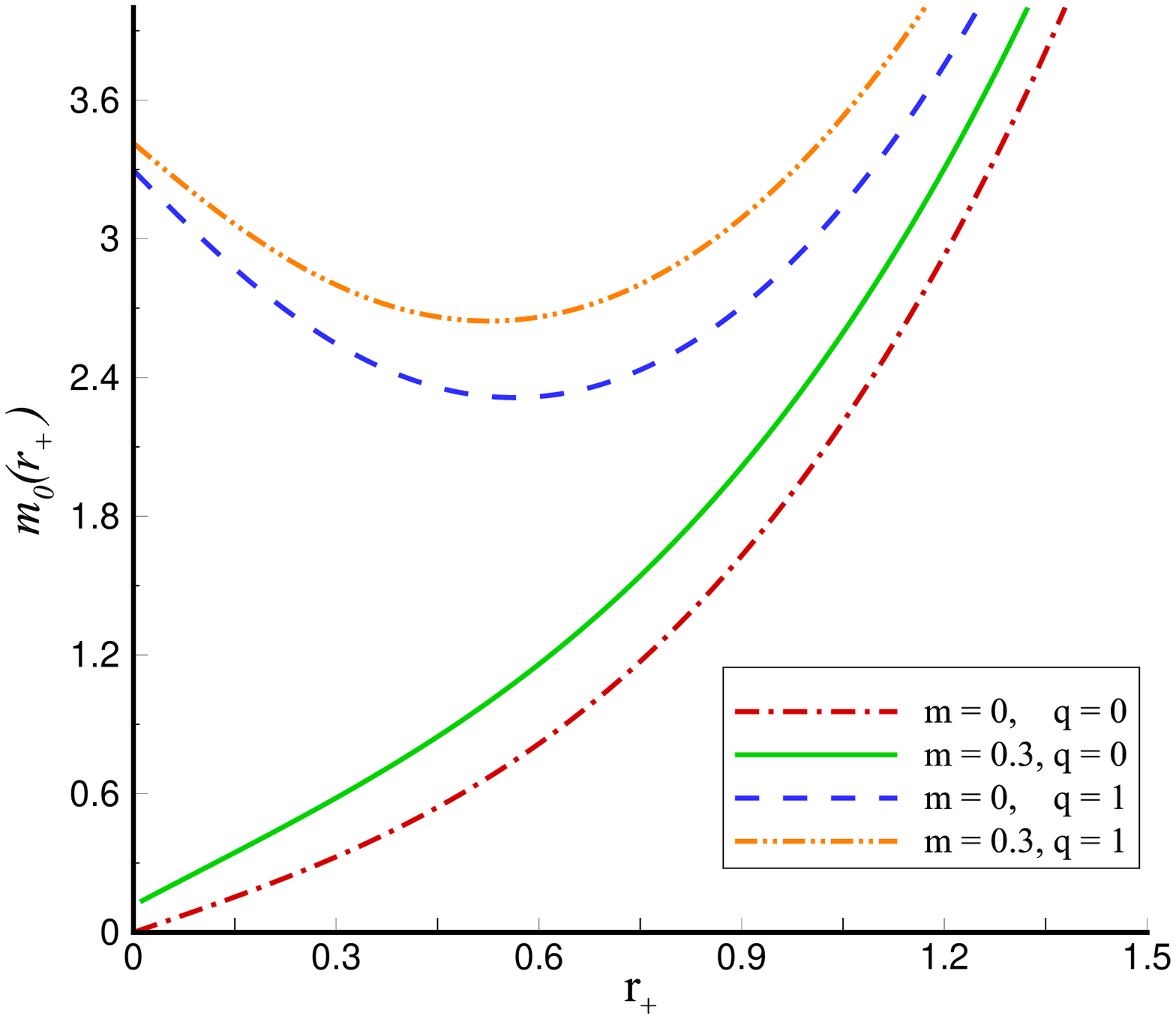}
        \caption{The behavior of $m_{0}(r_{+})$ versus $r_+$ for ~~~~~$\beta =1$,
         $n=3$, $c_{0}=-1$, $c_{1}=c_{2}=6$, $c_{3}=1.3$ and $c_{4}=0.2$.}
        \label{fig3}
    \end{minipage}
\hspace*{1cm} 
\begin{minipage}[t]{0.4\textwidth}
        \includegraphics
        [width=\linewidth,keepaspectratio=true]{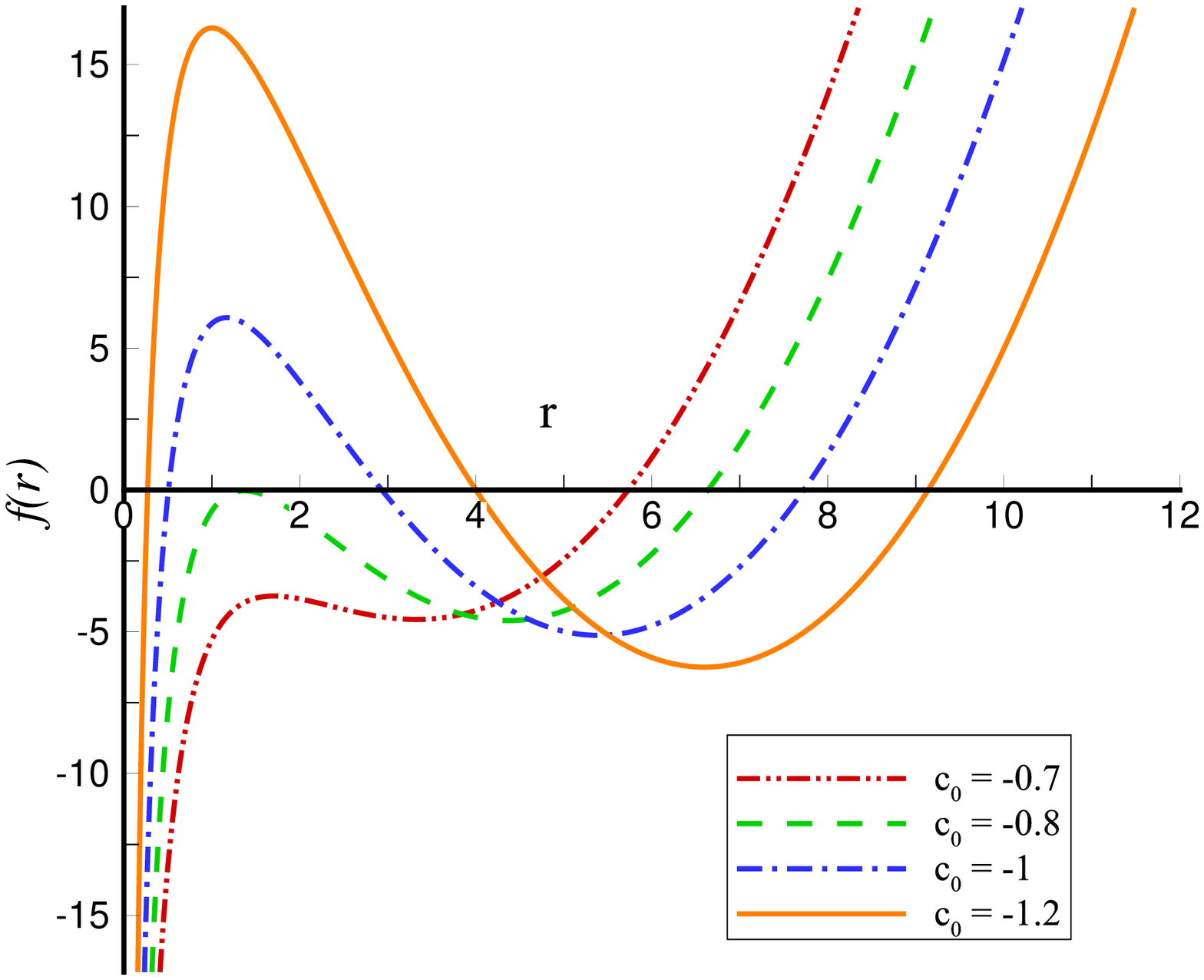}
        \caption{The behavior of $f(r)$ versus $r$ for $\protect\beta =5$, $%
            m_{0}=14.2$, $n=3$, $c_{1}=90$, $c_{2}=112$, $c_{3}=1.3$, $c_{4}=0.5$ and $%
            m=0.5$.}
        \label{fig2}
    \end{minipage}
\end{figure}
We also choose $m_{0}$ as a constant of integration which is related to
total mass of black hole. It is easy to obtain the parameter $m_{0}$ in (\ref%
{f(r)}) by using Eq. $f(r_{+})=0$. We find 
\begin{eqnarray}
m_{0} &=&kr_{+}^{n-2}-\frac{2r_{+}^{n}\Lambda }{n(n-1)}+\frac{8\beta
^{2}r_{+}^{n}}{n\left( n-1\right) }\left[ \ln \left( \frac{1+\Upsilon _{+}}{2%
}\right) +\frac{\left( 2n-1\right) \left( 1-\Upsilon _{+}\right) }{n}\right]
\notag \\
&&+\frac{8\left( n-1\right) q^{2}r_{+}^{-2n}}{n^{2}\left( n-2\right) }%
\mathbf{F}\left( \frac{1}{2},\frac{2-n}{2(-1+n)},\frac{4-3n}{2-2n},\frac{%
-q^{2}r_{+}^{2-2n}}{\beta ^{2}}\right)  \notag \\
&&+\frac{c_{0}m^{2}r_{+}^{n-1}}{n-1}\left( c_{1}+\frac{(n-1)c_{0}c_{2}}{r_{+}%
}+\frac{(n-1)(n-2)c_{0}^{2}c_{3}}{r_{+}^{2}}+\frac{%
(n-1)(n-2)(n-3)c_{0}^{3}c_{4}}{r_{+}^{3}}\right) ,  \label{m00}
\end{eqnarray}%
in which $\Upsilon _{+}=\Upsilon (r_{+})$ and $r_{+}$ is the radius of
horizon which is the largest root of $f(r_{+})=0$. As it is clear from Eq. (%
\ref{f(r)}), the position of $r_{+}$ completely depends on the metric
parameters. This fact is shown in Fig. \ref{fig1}. In this figure, we set $%
l=k=1$. Depending on the number of metric function's root(s), our solution
may be a black hole with two inner and outer horizons, an extreme black hole
or a naked singularity. As one can see from this figure, the number of
horizons, decreases by increasing the massive gravity parameter $m$ and
electric charge $q$. It is notable to mention that a suitable choice of
metric parameters may result a Schwarzschild-like black hole (Fig. \ref{fig1}%
). From Fig. \ref{fig3}, we observe that in the presence of massive gravity
and nonlinear electrodynamics, the mass parameter $m_{0}$ has non zero value
when $r_{+}\rightarrow 0$. Fig. \ref{fig3} also confirms the result of Fig. %
\ref{fig1b}. Comparing dash-dotted and dashed as well as solid and
dash-double-dotted curves in Fig. \ref{fig3} shows that by increasing $q$
when $m$ is fix, we could move from Schwarzschild-like black holes with just
one horizon to black holes with two horizons. One could also depict curves
to show the same behavior for $m$ (as shown before in Fig. \ref{fig1a}),
however we avoid this for economic reasons.

Fig. \ref{fig2} represents a significant behavior of the metric function $%
f(r)$; the existence of several roots! The green (dashed) curve shows two
roots where the smaller one is extreme and the blue (dash-dotted) curve
exhibits the existence of three roots. Also, as one decreases the $c_{0}$
parameter, the largest root moves to larger values. Similar result has been
reported in \cite{ref40}. It has been discussed that the physical reason
behind this property is anti-evaporation \cite{ref40}. Anti-evaporation is a
quantum effect which causes the size of black hole to increase.

The gauge potential $A_{t}$ can be calculated as 
\begin{align}
A_{t}\left( r\right) & =\int F_{rt}dr  \notag \\
& =\mu +\frac{2r^{n}\beta ^{2}}{nq}\left( \Upsilon -1\right) +\frac{2\left(
1-n\right) q^{2}r^{2-n}}{n\left( n-2\right) }\mathbf{F}\left(\frac{1}{2},%
\frac{n-2}{2(n-1)},\frac{4-3n}{2(1-n)},1-\Upsilon ^{2}\right).
\end{align}%
From holographic point of view, $\mu $ as a constant of integration is
chemical potential of quantum field that locates on the boundary. It can be
found by demanding the regularity condition on the horizon i.e. $%
A_{t}(r_{+})=0$

\begin{equation}
\mu =\frac{2r_{+}^{n}\beta ^{2}}{nq}\left( 1-\Upsilon _{+}\right) -\frac{%
2\left( 1-n\right) q^{2}r_{+}^{2-n}}{n\left( n-2\right) }\mathbf{F}\left(%
\frac{1}{2},\frac{n-2}{2(n-1)},\frac{4-3n}{2(1-n)},1-\Upsilon
_{+}^{2}\right).
\end{equation}%
We wind up this section by calculating the Hawking temperature on the event
horizon

\begin{eqnarray}
T &=&\frac{f^{\prime }(r_{+})}{4\pi }  \notag \\
&=&\frac{(n-2)k}{4\pi r_{+}}+\frac{m^{2}c_{0}}{4\pi r_{+}^{3}}\left(
c_{0}^{3}c_{4}(n-4)(n-3)(n-2)+c_{0}^{2}c_{3}(n-3)(n-2)r_{+}+c_{0}c_{2}(n-2)r_{+}^{2}+c_{1}r_{+}^{3}\right)
\notag \\
&&-\frac{r_{+}\Lambda }{2\pi (n-1)}+\frac{2r_{+}\beta ^{2}\left[ \left(
1-\Upsilon _{+}\right) +\ln \left( \frac{1}{2}\left( 1+\Upsilon _{+}\right)
\right) \right] }{\pi (n-1)}.  \label{temp}
\end{eqnarray}%
In the next section, we will study the thermodynamics of our black hole
solutions.

\section{THERMODYNAMICS OF MASSIVE GRAVITY SOLUTIONS \label{Therm}}

In order to investigate the thermodynamics of our black hole solutions, we
have to calculate some thermodynamical quantities. We first obtain the
entropy of the black holes. Based on \cite{ref39}, this thermodynamic
quantity is equal to one-quarter of the horizon area 
\begin{equation}
S=\frac{r_{+}^{n-1}}{4}.  \label{entropy}
\end{equation}%
The Gauss law allows us to compute electric charge per unit volume $V_{n-1}$
using electromagnetic flux at infinity,%
\begin{equation}
Q=\frac{\,{1}}{4\pi }\int r^{n-1}\left( 1+\frac{\mathcal{F}}{8\beta ^{2}}%
\right) ^{-1}F_{\mu \nu }n^{\mu }u^{\nu }dr,
\end{equation}%
in which the unit spacelike and timelike normals to the hypersurface of
radius $r$ are respectively $n^{\mu }=\left( \sqrt{-g_{tt}}\right)
^{-1}dt=\left( \sqrt{f\left( r\right) }\right) ^{-1}dt$ and $u^{\nu }=\left( 
\sqrt{g_{rr}}\right) ^{-1}dr=\sqrt{f(r)}dr$. Finally, the electric charge
per unit volume $V_{n-1}$ of black hole could be obtained as%
\begin{equation}
Q=\frac{\,q}{4\pi }.  \label{charge}
\end{equation}%
Electric potential can measure at infinity with respect to the horizon using
the following definition 
\begin{equation}
U=A_{\mu }\chi ^{\mu }\left\vert _{r\rightarrow \infty }-A_{\mu }\chi ^{\mu
}\right\vert _{r=r_{+}},  \label{Pot}
\end{equation}%
where $\chi =\partial _{t}$ is the null generator of the horizon. So it is
easy to show that electric potential is equal to chemical potential 
\begin{equation}
U=\mu .  \label{Poten}
\end{equation}%
The mass of our charged solution using the Hamiltonian approach can be found
as%
\begin{equation}
M=\frac{(n-1)m_{0}}{16\pi }.  \label{Mass}
\end{equation}

\begin{figure}[t]
\centering\subfigure[~$m=5$]{\includegraphics[scale=0.3]{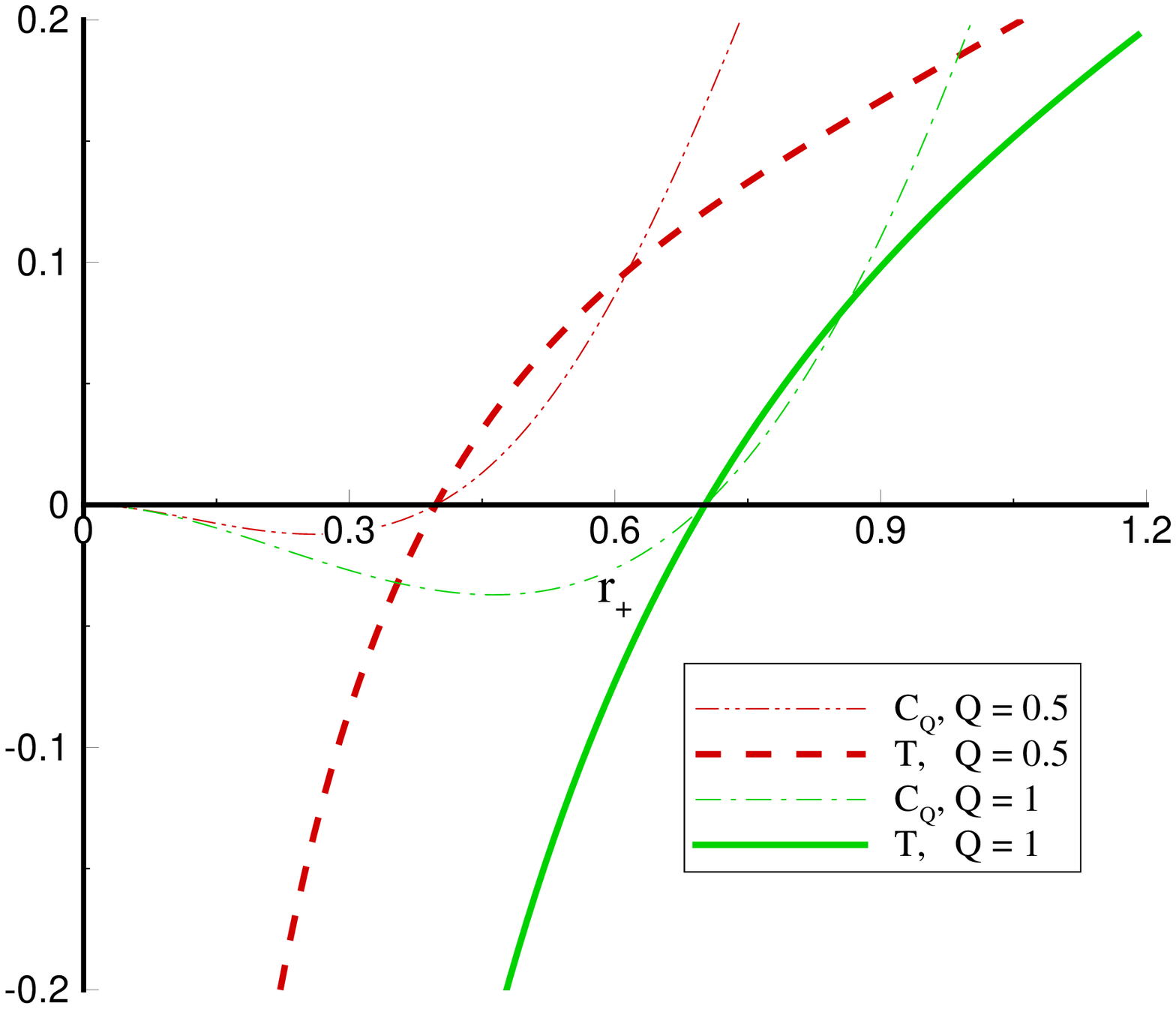}\label{fig4a}} 
\hspace*{0.5cm} 
\subfigure[~$q=0.5$]{\includegraphics[scale=0.3]{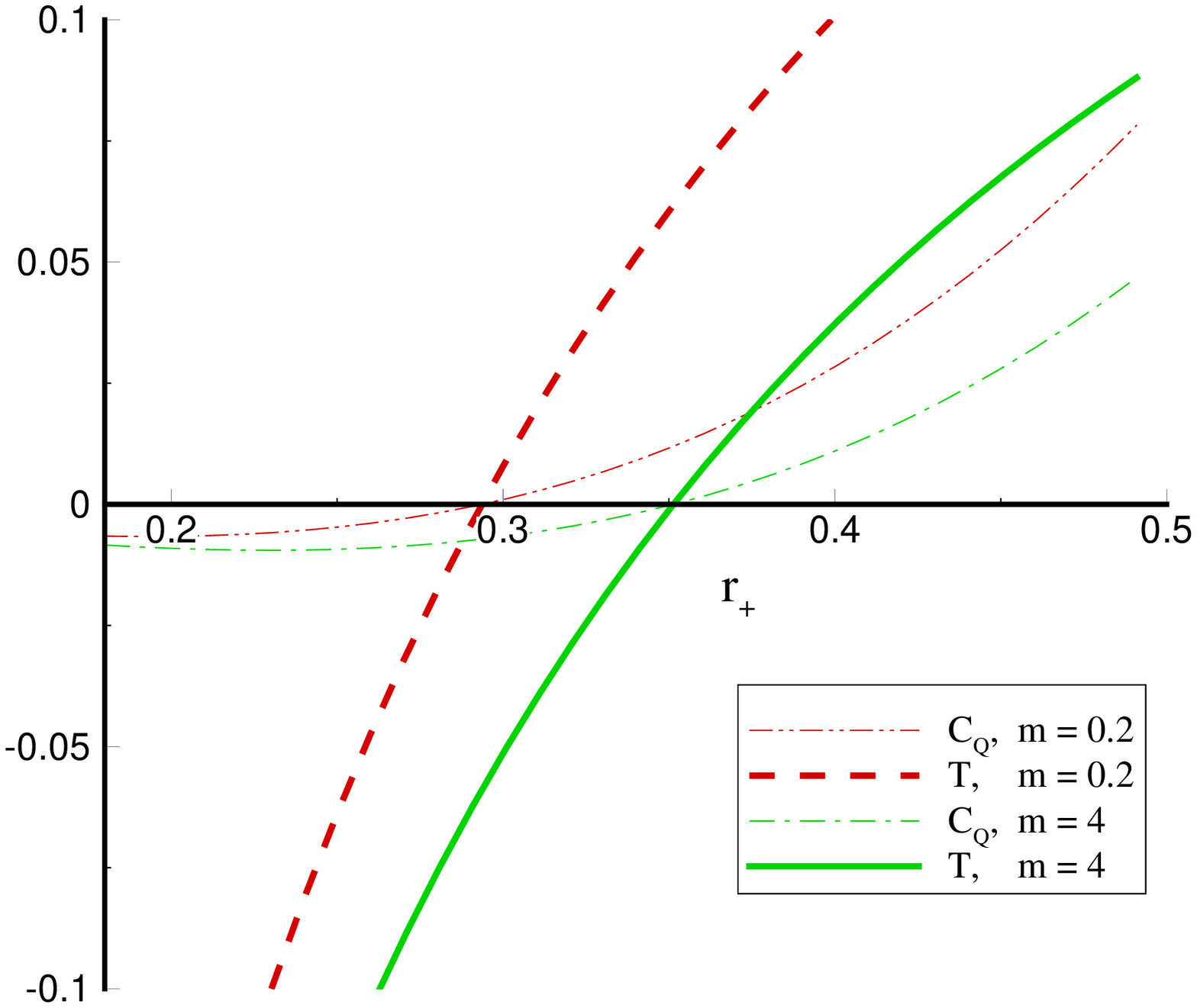}
\label{fig4b}}
\caption{The behavior of $C_{Q}$ and $T$ versus $r_{+}$ for $\protect\beta %
=1 $, $m_{0}=0.2$, $n=3$, $c_{0}=-0.2$, $c_{1}=0.3$, $c_{2}=0.1$, $c_{3}=1.3$
and $c_{4}=5$.}
\label{fig4}
\end{figure}

Using (\ref{m00}), (\ref{entropy}), (\ref{charge}) and (\ref{Mass}), it is a
matter of calculation to show that the mass function is

\begin{align}
M(S,Q)=& -\frac{\Lambda \,(4S)^{\frac{n}{n-1}}}{8n\pi }+\frac{m^{2}c_{0}}{%
16\pi }\left[ (n-1)(n-2)(4S)^{\frac{n-3}{n-1}}\left( c_{3}+(4S)^{\frac{-1}{%
n-1}}(n-3)c_{4}\right) c_{0}^{3}+(4S)^{\frac{n-2}{n-1}%
}(n-1)c_{2}c_{0}+4Sc_{1}\right]  \notag \\
& +\frac{\beta ^{2}}{2n\pi }\ln \left[ \frac{1}{2}+\frac{1}{2}\sqrt{1+\frac{%
Q^{2}\pi ^{2}}{\beta ^{2}S^{2}}\,}\right] (4S)^{\frac{n}{n-1}}+\frac{%
8(n-1)^{2}\,Q^{2}\pi }{n^{2}(n-2)}\mathbf{F}\left( \frac{1}{2},\frac{n-2}{%
2n-2},\frac{3n-4}{2-2n},-\frac{Q^{2}\pi ^{2}}{\beta ^{2}S^{2}}\right) (4S)^{%
\frac{2-n}{n-1}}  \notag \\
& +\frac{1}{16n^{2}\pi }\left[ 8\beta ^{2}(2n-1)+k\,n^{2}(n-1)(4S)^{\frac{-2%
}{n-1}}-8\beta ^{2}(2n-1)\sqrt{1+\frac{Q^{2}\pi ^{2}}{\beta ^{2}S^{2}}}\,%
\right] (4S)^{\frac{n}{n-1}}
\end{align}%
In order to check the validity of the first law of thermodynamics for our
solutions, we consider entropy $S$ and electric charge $Q$ as a complete set
of extensive quantities for mass. Also, we define temperature $T$ and
electric potential $U$ as their conjugate intensive quantities, respectively:%
\begin{equation}
T=\left( \frac{\partial M(S,Q)}{\partial S}\right) _{Q}\text{ \ \ \ \ and \
\ \ \ }U=\left( \frac{\partial M(S,Q)}{\partial Q}\right) _{S}.
\label{intqua}
\end{equation}%
Our numerical calculations show that, above quantities satisfy the first law
of black hole thermodynamics: 
\begin{equation}
dM=TdS+UdQ.  \label{TFL}
\end{equation}%
In following section, we will investigate the thermal stability of obtained
solutions.

\section{Thermal Stability \label{Stability}}

\begin{figure}[t]
\centering
\subfigure[~$n=3$, $c_1=4$, $c_2=10$, $c_3=13$, $c_4=0.5$, $q=1$,
    $m=0.5$]{\includegraphics[scale=0.3]{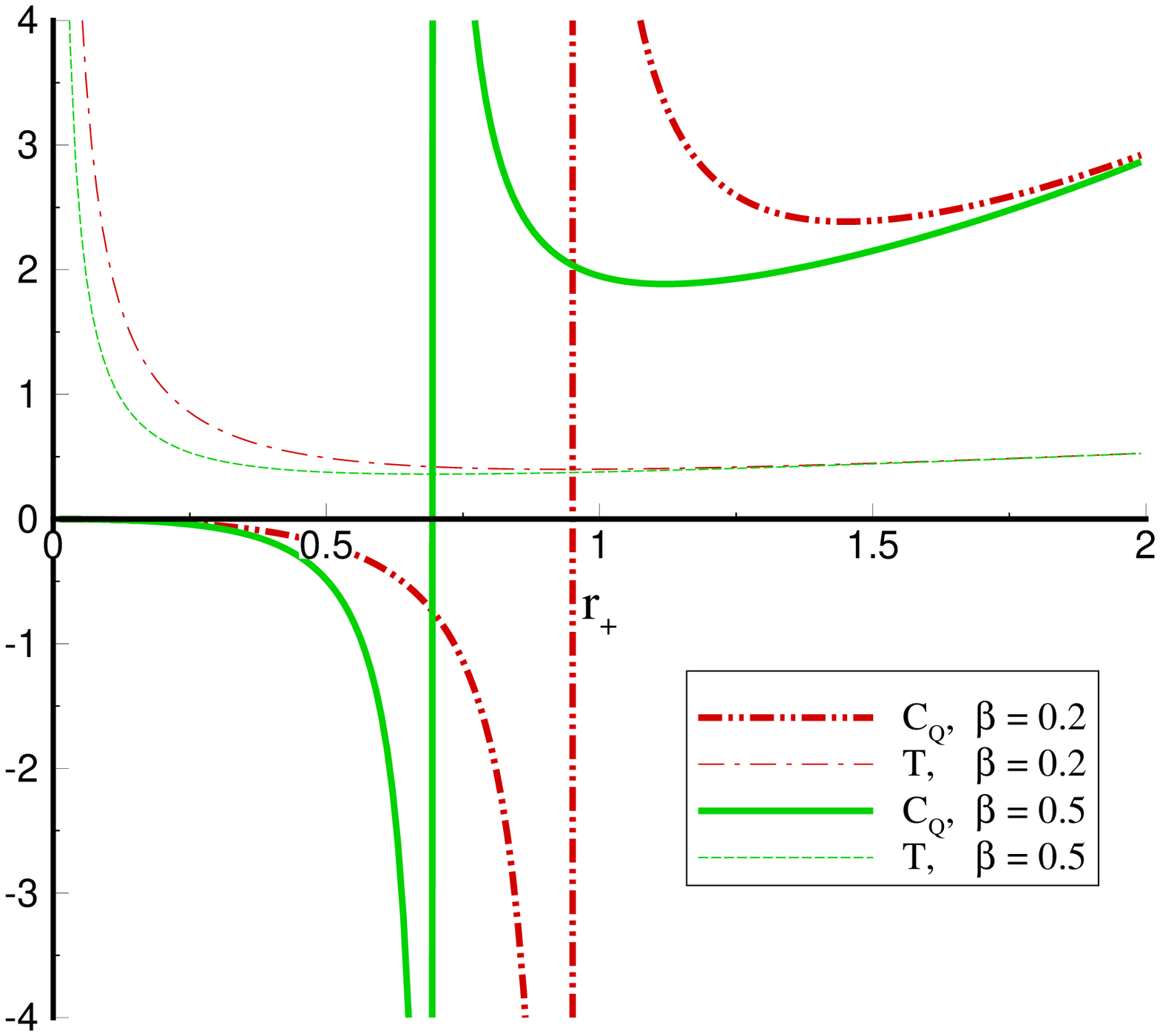}\label{fig6}} \hspace*{.5cm} 
\subfigure[~$\beta=0.4$, $c_1=3$, $c_2=2$, $c_3=2$, $c_4=1$, $q=0.5$,
$m=0.8$]{\includegraphics[scale=0.3]{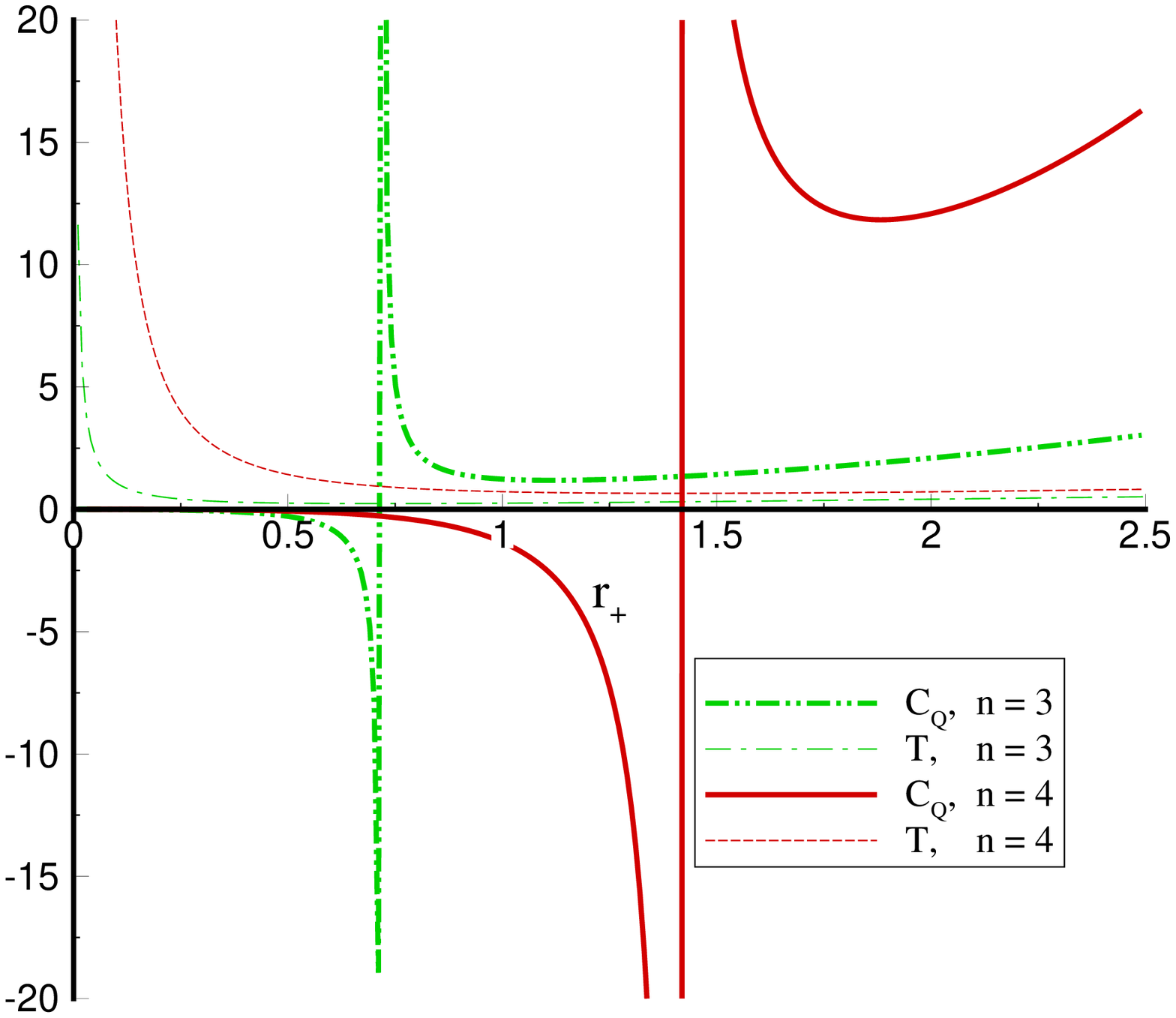}\label{fig7}} 
\caption{The behavior of $C_{Q}$ and $T$ versus $r_{+}$ for $m_{0}=2$ and $%
c_{0}=-1$.}
\label{fig5}
\end{figure}
\begin{figure}[t]
\centering\includegraphics[scale=0.3]{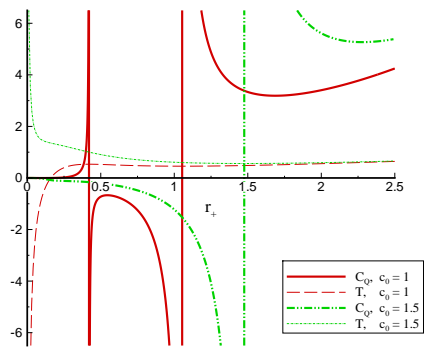}
\caption{The behavior of $C_{Q}$ and $T$ versus $r_{+}$ for $\protect\beta %
=3 $, $m_{0}=2$, $n=3$, $c_{1}=1$, $c_{2}=3$, $c_{3}=1$, $c_{4}=-7$, $q=0.5$
and $m=1$.}
\label{fig8}
\end{figure}
This section is devoted to study the thermal stability of the obtained
solutions. One can consider the stability of a black hole in both canonical
and grand canonical ensembles. In canonical ensemble, positivity of the heat
capacity together with the positive values for the temperature guaranties
thermal stability of the solutions. Heat capacity is defined as 
\begin{equation}
C_{Q}=T\left( \frac{\partial S}{\partial T}\right) _{Q}=\left( \frac{%
\partial M}{\partial S}\right) _{Q}\left( \frac{\partial ^{2}M}{\partial
S^{2}}\right) _{Q}^{-1}.
\end{equation}%
In case of unstable black hole, a phase transition occur to get stable
state. We desire to investigate the phase transition point by finding the
roots and divergencies of $C_{Q}$. The roots of heat capacity which are
equivalent to zero values of temperature, determine a thermal transition
between un-physical ($T<0$) and physical ($T>0$) state of the black hole.
Beside, we recall that based on the thermal physics, when the heat capacity
diverges, a second order phase transition will occur.

Since it is not easy to determine the roots and divergencies of the heat
capacity analytically, we study this by using the figures. The behaviors of $%
C_{Q}$ in terms of $r_{+}$ for different sets of parameters have been
depicted in Figs. \ref{fig4}-\ref{fig8}. It is obvious that the stability
depends on the metric parameters. By choosing a proper set of metric
parameters, Fig. \ref{fig4} specifies that by increasing the electric charge 
$Q$ (massive parameter $m$), the transition between un-physical and physical
states will occur for larger black holes.

Fig. \ref{fig5} illustrates a second order phase transition of our solutions
for some specific values of the model parameters. The transition is between
two physical states since the temperature is positive. In Fig. \ref{fig5},
we observe that for smaller black holes $C_{Q}$ is negative, so they are not
stable and transition allows us to have larger stable black holes. In
addition, as we decrease the nonlinearity of electrodynamics (larger $\beta $%
), the transition happens for smaller black holes (Fig. \ref{fig6}). As Fig. %
\ref{fig7} shows, for higher dimensions phase transition occurs for larger
horizon radii.

Changing massive model parameters causes rich phase transitions (Fig. \ref%
{fig8}). The solid curve in Fig. \ref{fig8} shows two transitions. One from
stable small black hole to unstable middle black hole and another from
unstable middle black hole to stable large black hole. These transitions
represent that depending on the metric parameters, the middle black holes
may not be allowed. This fact is clear from dash-double-dotted curve for
which the transition is just from small to large black holes. 
\begin{figure}[t]
\centering
\subfigure[~$c_0=-2$, $c_2=3$, $c_4=0.5$,
    $n=3$]{\includegraphics[scale=0.3]{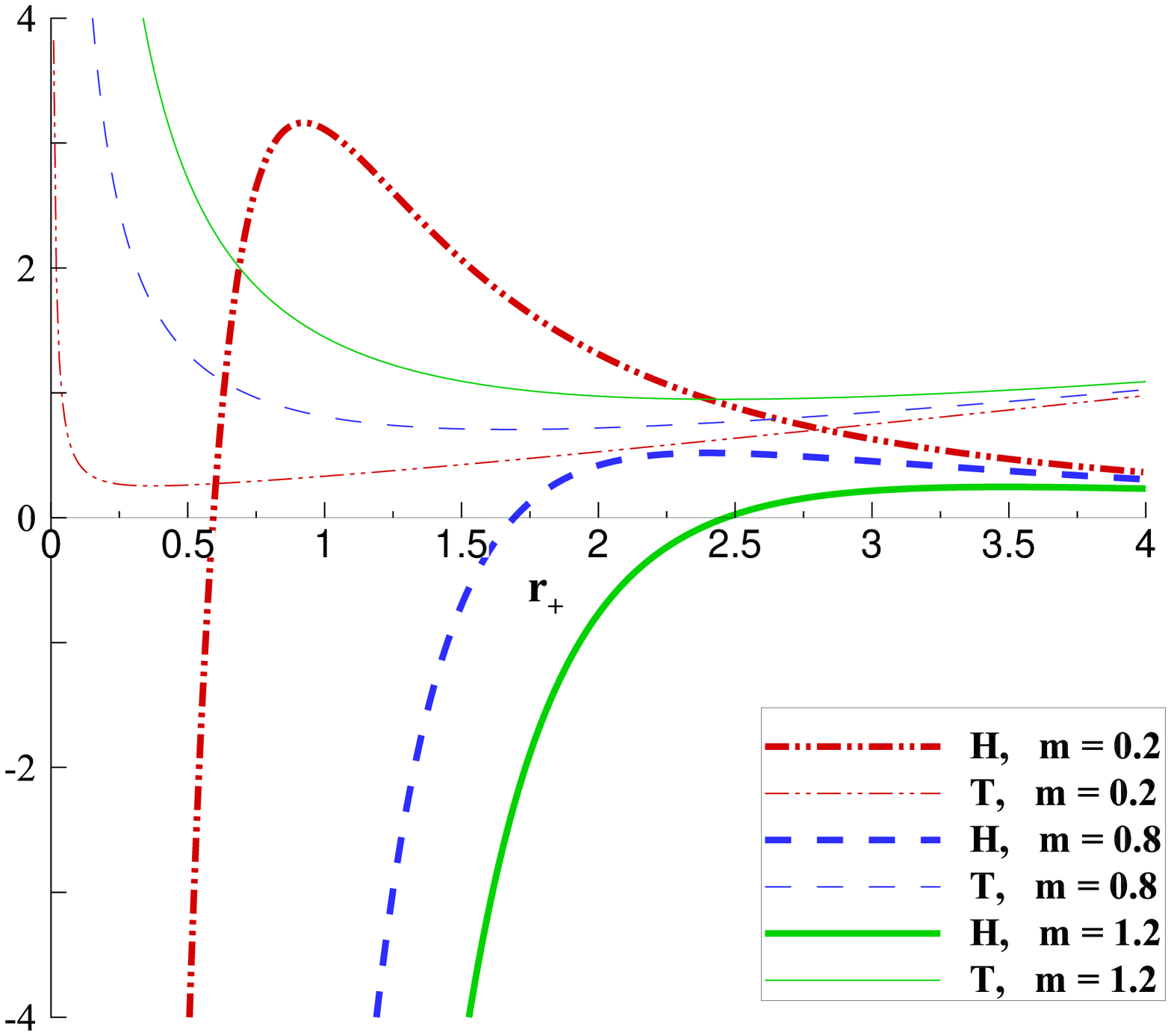}\label{fig9a}} \hspace*{.5cm} 
\subfigure[~$c_0=-0.8$, $c_2=1$, $c_4=1$,
    $m=1.2$]{\includegraphics[scale=0.3]{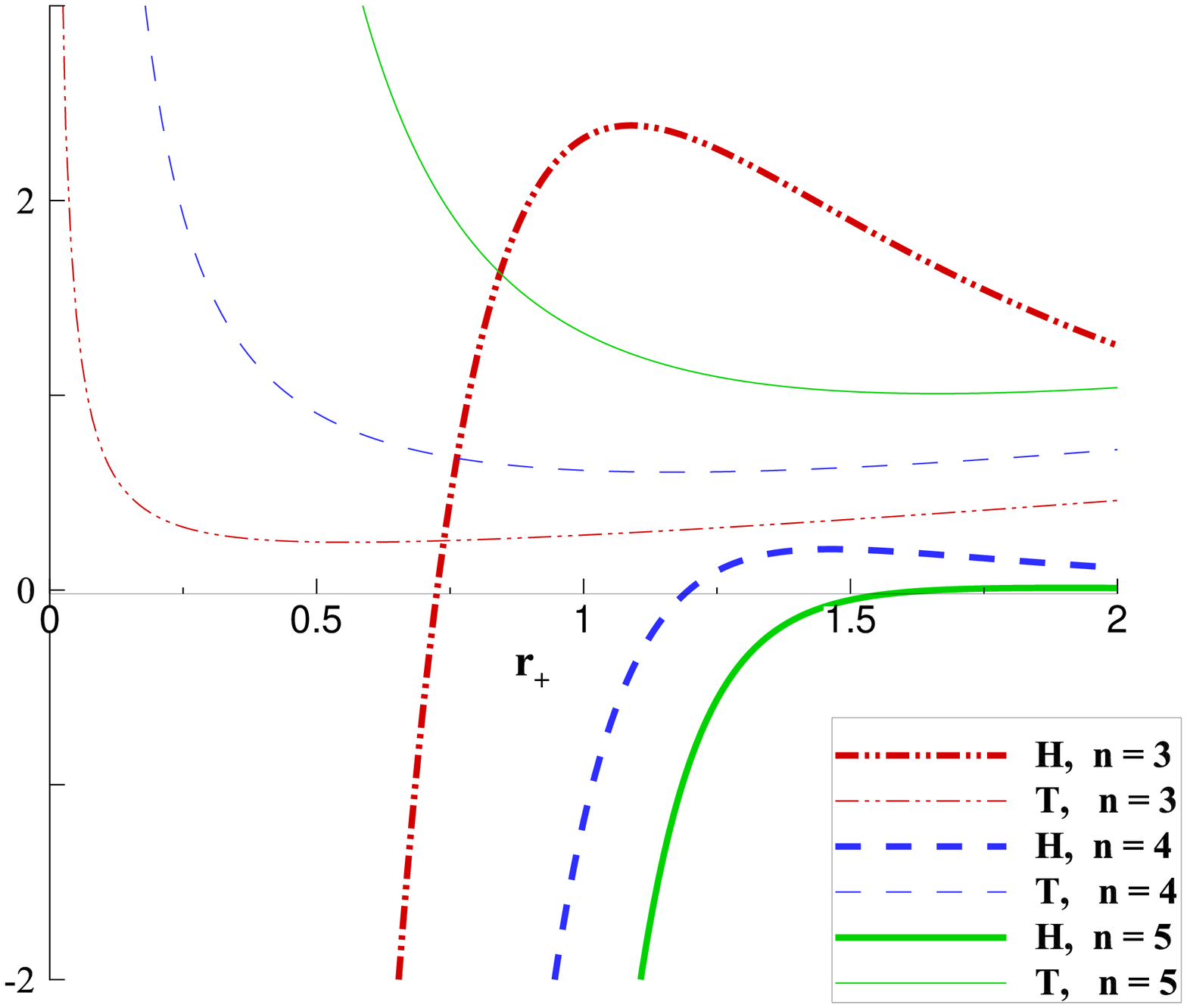}\label{fig9b}} 
\caption{The behavior of $\mathbf{H}$ and $T$ versus $r_{+}$ for $\protect%
\beta =0.5$ , $m_{0}=2$, $c_{1}=1$, $c_{3}=1$ and $Q=0.5$.}
\label{fig9}
\end{figure}
\begin{figure}[t]
\centering
\subfigure[~$Q=1.5$, $m_0=1$, $c_2=3$, $c_3=13$, $c_4=-1$,
    $m=1$]{\includegraphics[scale=0.3]{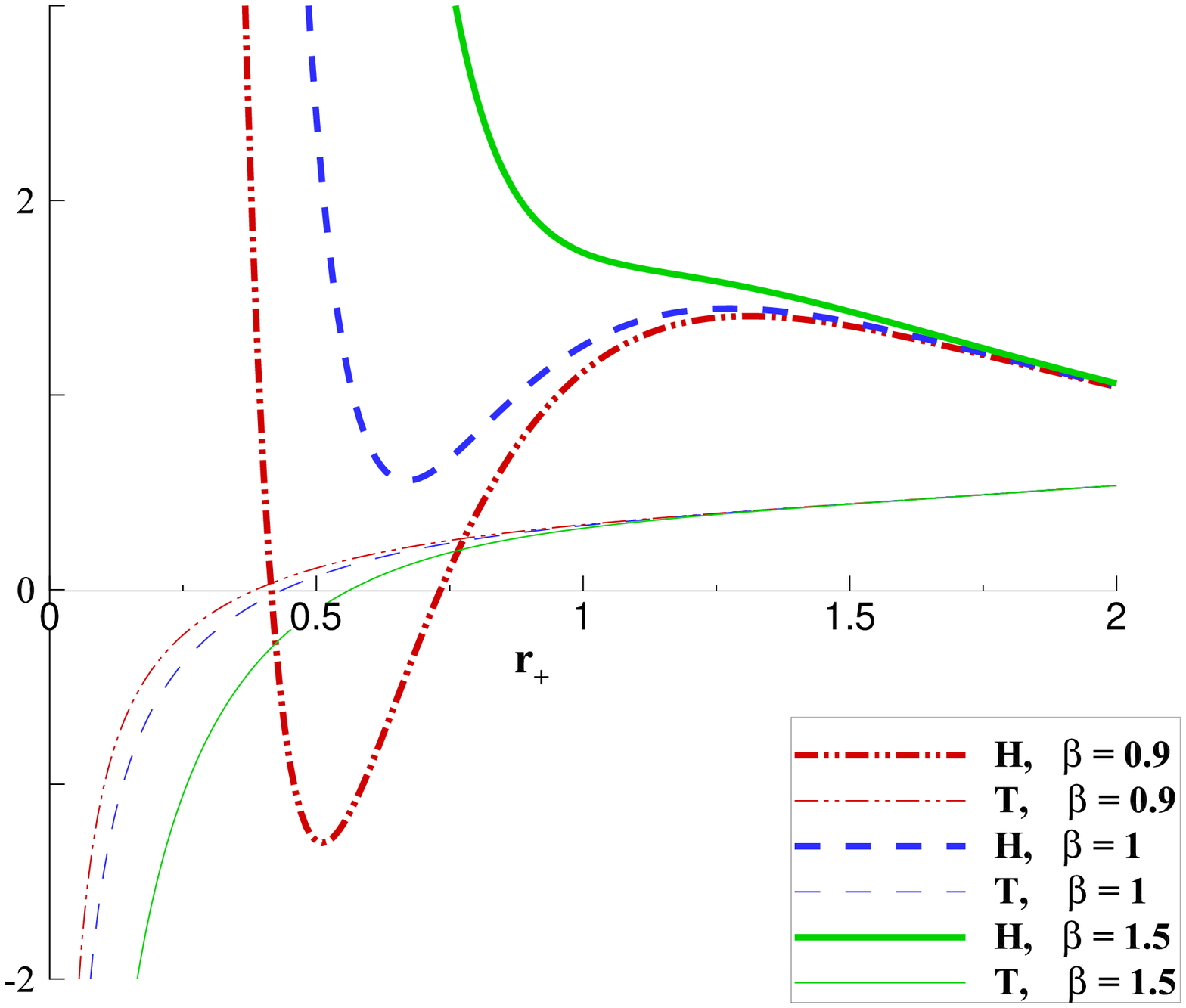}\label{fig10a}} \hspace*{.5cm} 
\subfigure[~$\beta=2$, $m_0=2$, $c_2=1$, $c_3=2$, $c_4=1$,
    $m=1.2$]{\includegraphics[scale=0.3]{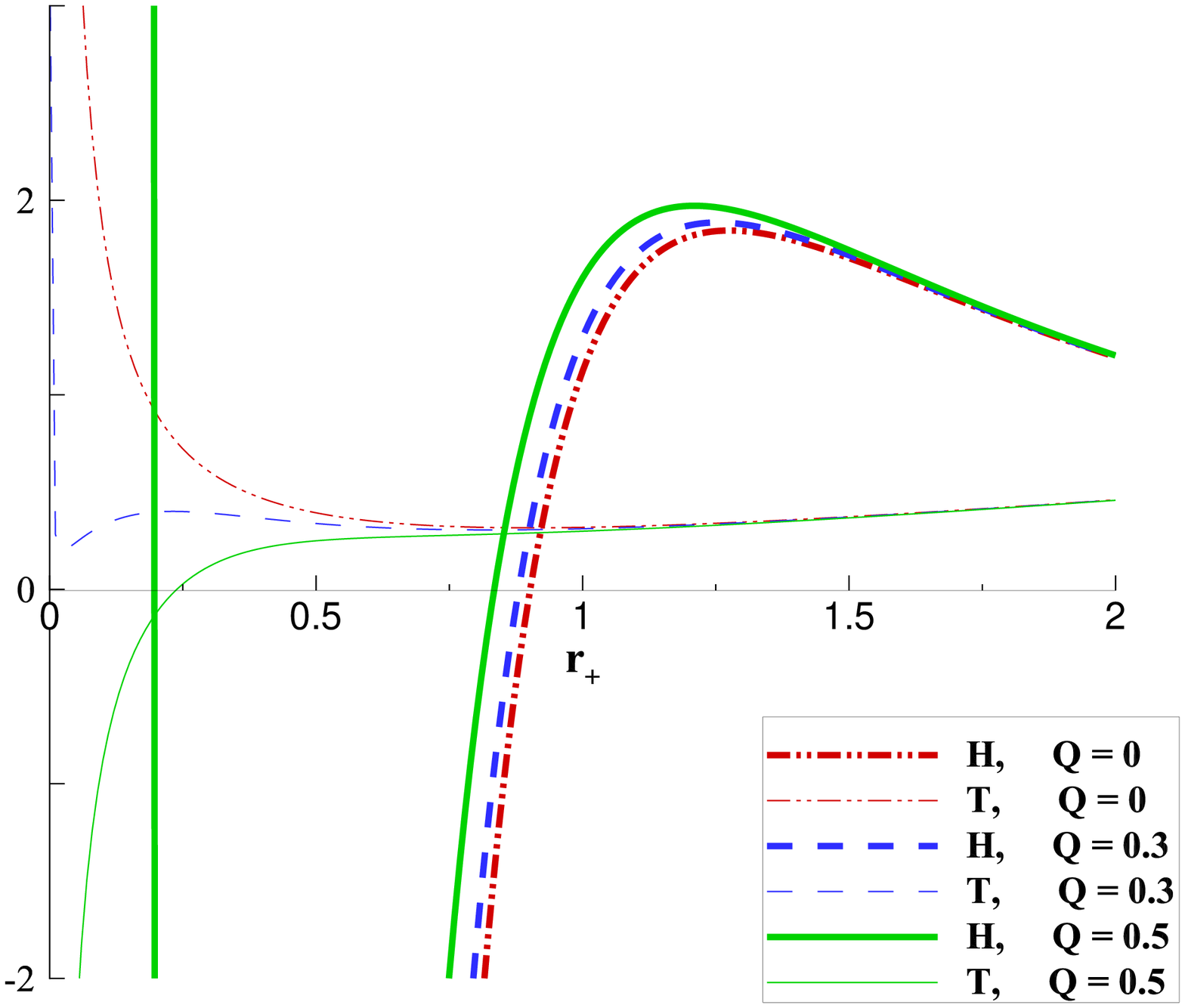}\label{fig10b}} 
\caption{The behavior of $\mathbf{H}$ and $T$ versus $r_{+}$ for $n=3$, $%
c_{0}=-1$, $c_{1}=1$ and $c_{3}=1$.}
\label{fig10}
\end{figure}

Now, we turn to study the stability in the grand canonical ensemble.
Basically, the local stability can be carried out by finding the determinant
of the Hessian matrix $\mathbf{H}$ with respect to its extensive variables, $%
X_{i}$'s. In the canonical ensemble, $Q$ is a fixed parameter so the
positivity of heat capacity ($C_{Q}$) is sufficient to ensure the local
thermal stability. However, in grand canonical ensemble, the number of
extensive parameters depends on the theory, and we have to check the sign of 
$\mathbf{H}$ to search for thermal stability. In our case, the mass is a
function of extensive variables entropy $S$ and charge $Q$. So, Hessian
matrix defines as

\begin{equation}
\mathbf{H}_{X_{i},X_{j}}^{M}=\frac{\partial ^{2}M}{\partial X_{i}\partial
X_{J}}=\left[ 
\begin{array}{cc}
\frac{\partial ^{2}M}{\partial S^{2}} & \frac{\partial ^{2}M}{\partial
S\partial Q} \\ 
\frac{\partial ^{2}M}{\partial S\partial Q} & \frac{\partial ^{2}M}{\partial
Q^{2}}%
\end{array}%
\right].
\end{equation}
It is not easy to investigate the thermal stability analytically, so we turn
to the figures again. The behavior of $\mathbf{H}$ in terms of $r_{+}$ for
different sets of model parameters has been plotted in Fig. \ref{fig9} and %
\ref{fig10}. By fixing other metric parameters, Fig.\ref{fig9a} shows a
minimum value for horizon radius that for values under which black holes are
unstable. This minimum value grows with increasing the mass parameter of
massive gravity. The same behavior could be seen as well in terms of
dimension $n$ (Fig. \ref{fig9b}). We emphasize that in this range the
temperature is positive.

Changing nonlinear parameter $\beta $, may create two roots for $\mathbf{H}$
(Fig. \ref{fig10}). But since the temperature is negative for smaller root,
the only important limit for allowed horizon radius value depends on the
larger root. As one can see from Fig. \ref{fig10}, the horizon of stable
black hole should be larger than larger root. It is notable to mention that
by increasing $\beta $ this limit is removed. We observe from Fig.\ref%
{fig10b} that increasing electric charge $Q$ causes two roots for $\mathbf{H}
$. We do not care to the smaller root since it places in the negative range
of temperature. Furthermore, the larger root grows for smaller electric
charges, so the stable black holes have larger horizons.

\section{Critical behavior and Reentrant phase transition \label{Reentrant}}

\begin{figure}[t]
\centering\includegraphics[scale=0.7]{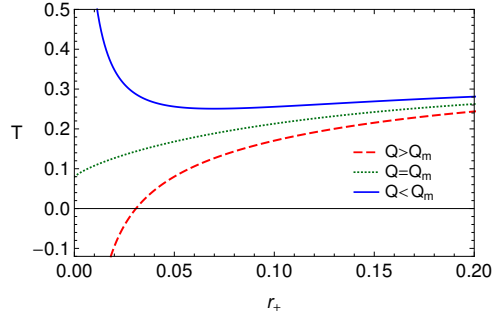}
\caption{$T-r_{+}$ diagram of ($3+1$)-dimensional logarithmic charged
massive black hole. We have set $c_{0}=1$, $c_{1}=1$, $c_{2}=-0.1$, $m=1$, $%
l=1$ and $\protect\beta =1$.}
\label{figure13}
\end{figure}

\begin{figure}[t]
\centering \subfigure[]{\includegraphics[scale=0.45]{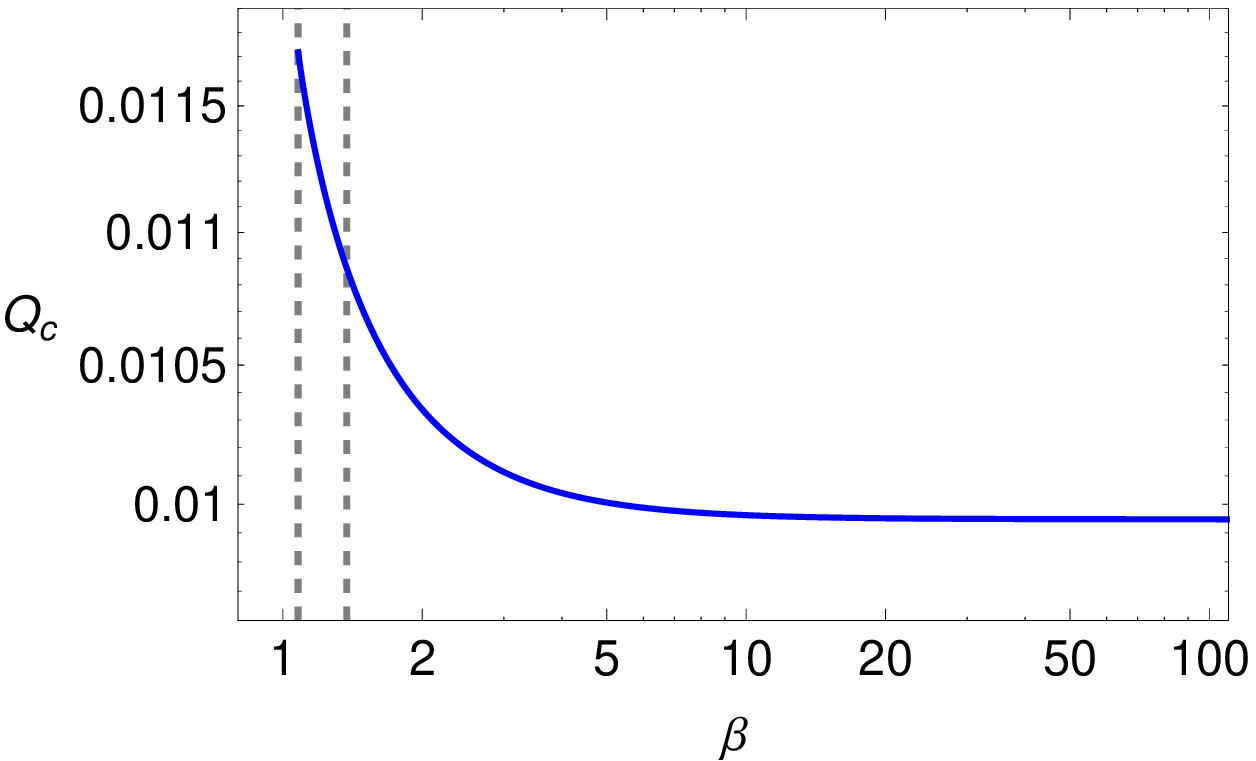}} \hspace*{0.1cm} %
\subfigure[]{\includegraphics[scale=0.45]{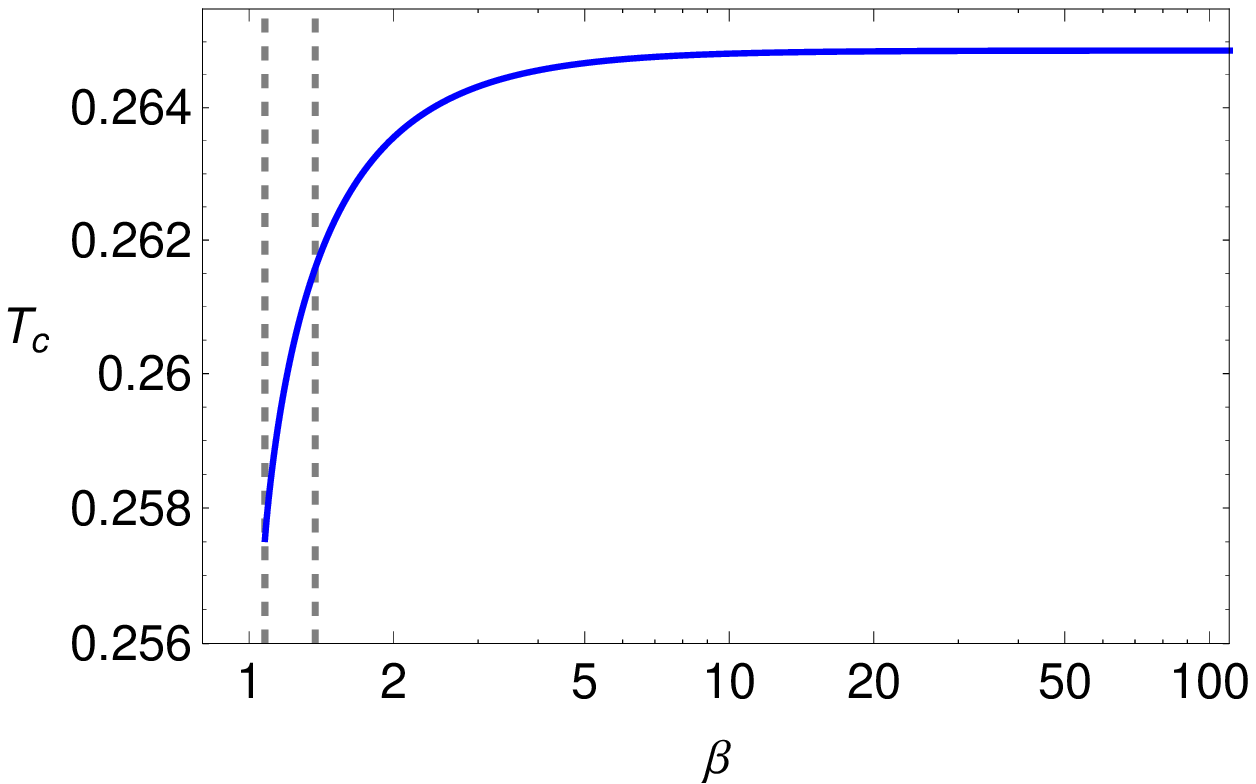}}\hspace*{0.1cm} %
\subfigure[]{\includegraphics[scale=0.45]{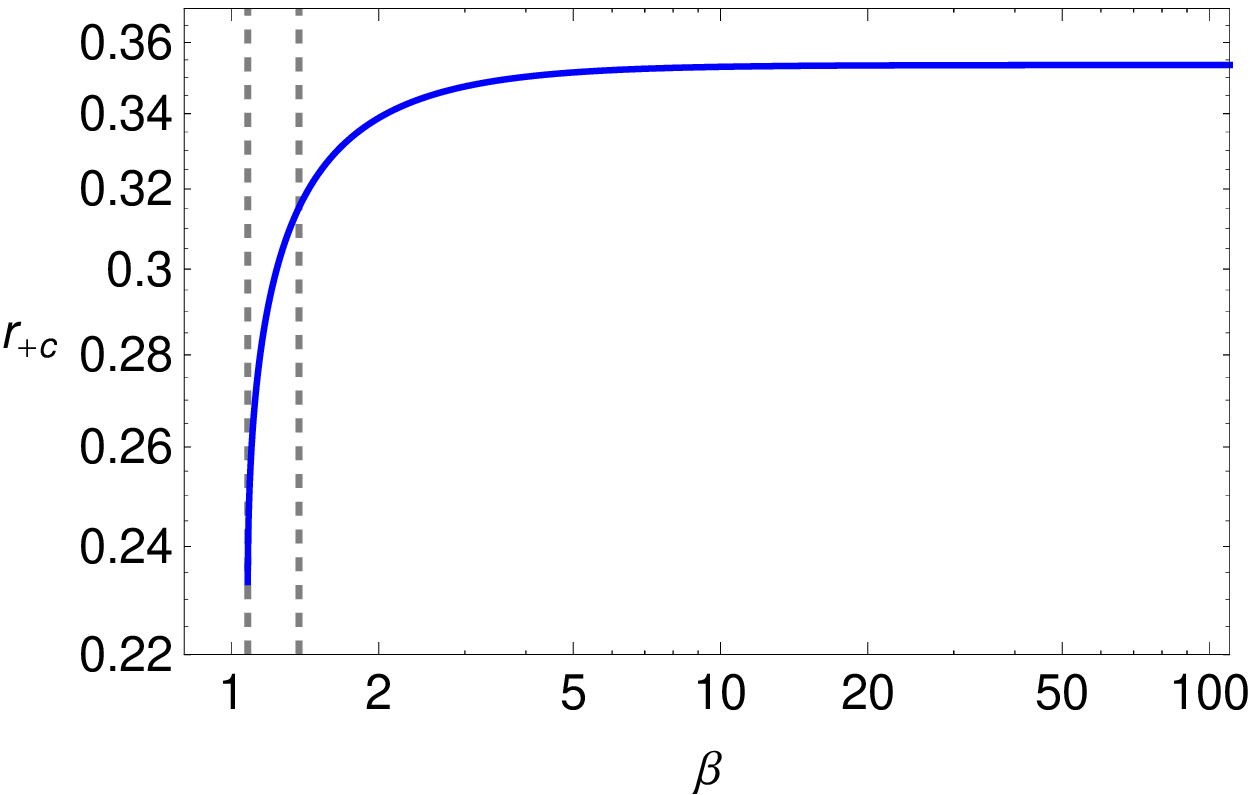}}
\caption{{}The behaviors of $Q_{c}$, $T_{c}$ and $r_{+c}$ vs $\protect\beta $
with $c_{0}=0.5$, $c_{1}=1$, $c_{2}=-1$, $m=1$ and $l=1$. Note that the
horizontal axes are logarithmic.}
\label{figure14}
\end{figure}

\begin{figure}[t]
\begin{minipage}[t]{0.4\textwidth}
        \includegraphics
        [width=\linewidth,keepaspectratio=true]{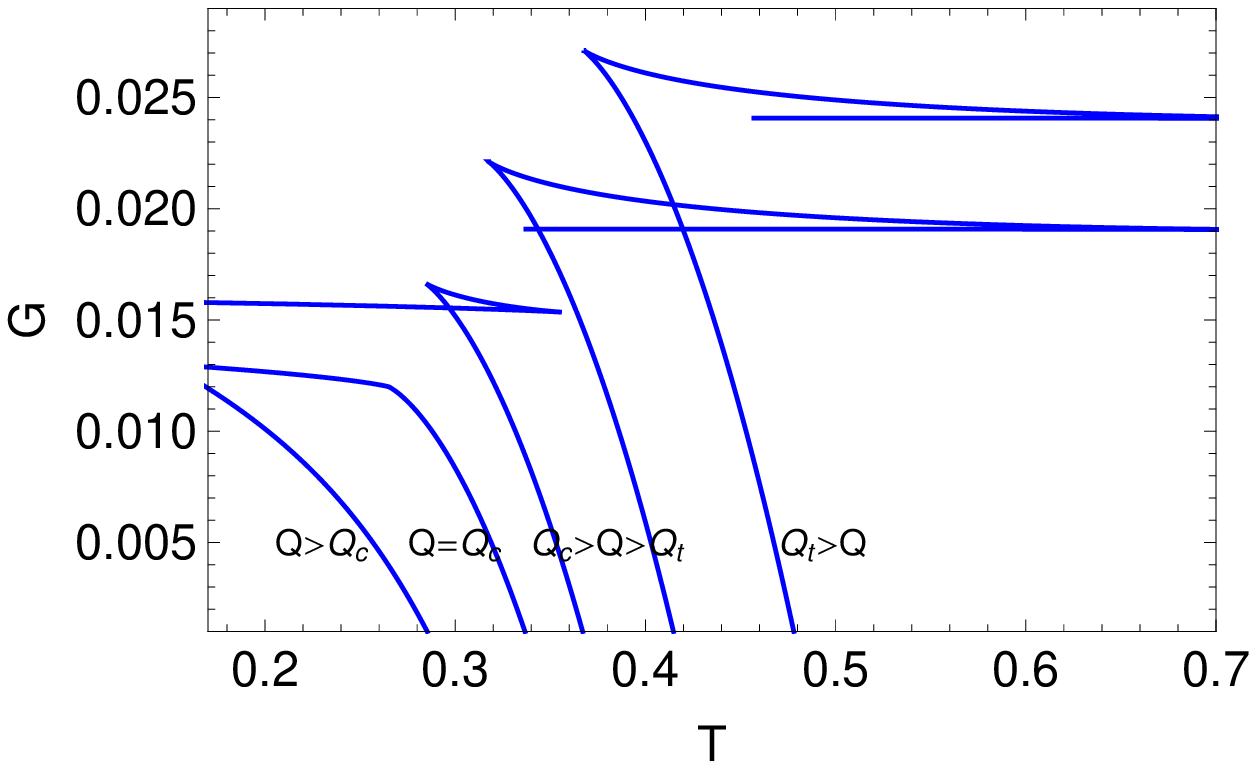}
        \caption{Gibbs free energy as a function of temperature for RN-type black
holes with different values of charge.
We fix $c_{0}=0.5$, $c_{1}=1$, $c_{2}=-1$, $m=1$, $l=1$ and $%
\protect\beta =5$.
Note that curves are shifted for clarity.}
        \label{figure15}
    \end{minipage}
\hspace*{1cm} 
\begin{minipage}[t]{0.4\textwidth}
        \includegraphics
        [width=\linewidth,keepaspectratio=true]{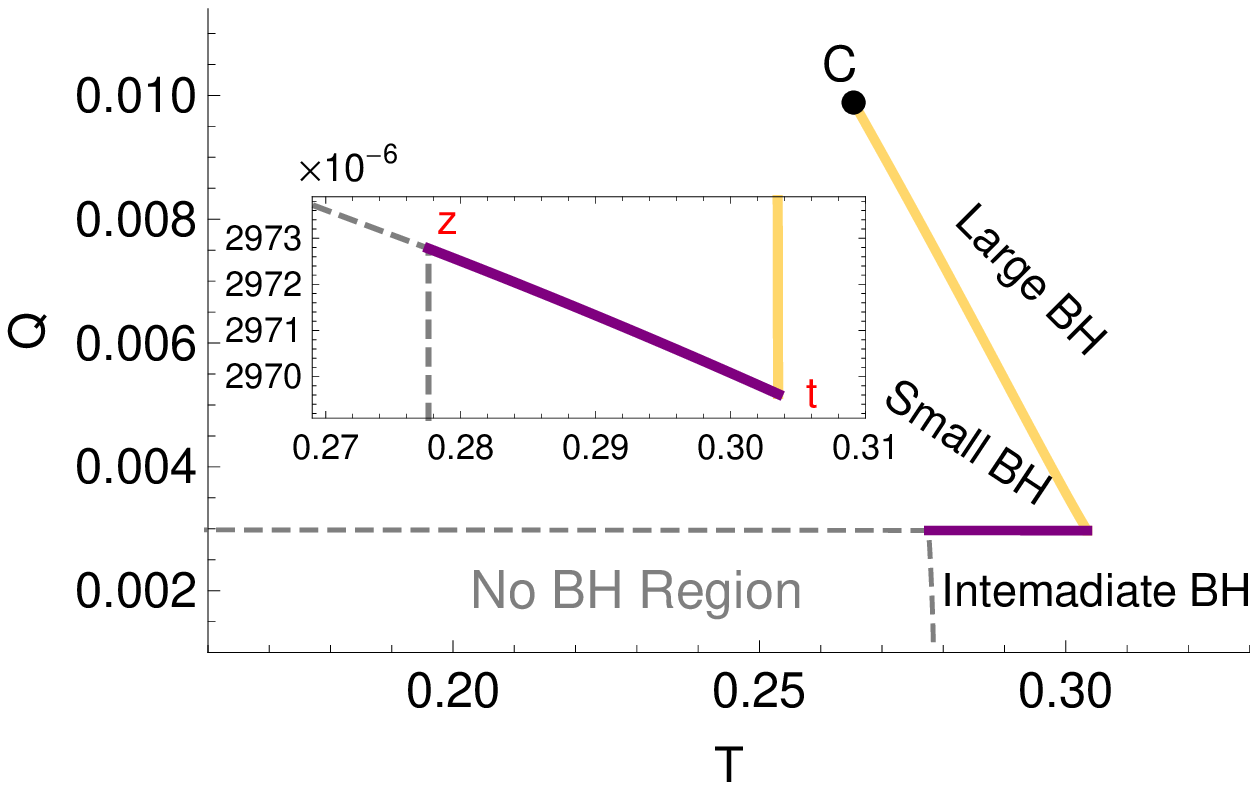}
        \caption{Phase diagram corresponding to RN-type black holes with $c_{0}=0.5$%
, $c_{1}=1$, $c_{2}=-1$, $m=1$, $l=1$ and $\protect\beta =5$. Note that No
BH region corresponds to no black hole solution for region of parameter
space.}
        \label{figure16}
    \end{minipage}
\end{figure}

In this section, we first explore critical behavior of ($3+1$)-dimensional
logarithmic charged massive black hole solutions with spherical event
horizon ($k=1$) by investigating the behavior of the specific heat at
constant charge, $C_{Q}$. Then, by studying Gibbs free energy $G$, we seek
for different phase transitions of systems.

The specific heat at constant charge is defined as%
\begin{equation}
C_{Q}=T\left( \frac{dS}{dT}\right) _{Q}.  \label{CQ}
\end{equation}%
The sign of $C_{Q}$ indicates the local thermodynamic stability/instability
of the system. The positive (negative) sign of this quantity shows the local
thermodynamic stability (instability). Note that Eq. (\ref{CQ}) is also
calculated at fixed cosmological constant and parameter of nonlinearity $%
\beta $. In Fig. \ref{figure13}, we plot the behavior of temperature with
respect to event horizon radius for different values of the charge. The sign
of slope in $T-r_{+}$ curves in allowed regions with positive temperature
determines the sign of $C_{Q}$. Expanding the Hawking temperature (\ref{temp}%
) for small $r_{+}$, we receive%
\begin{equation}
T=\frac{4\beta }{r_{+}}\left( Q_{m}-Q\right) +\frac{c_{0}c_{1}m^{2}}{4\pi }+%
\frac{r_{+}\left( 3+4l^{2}\beta ^{2}\left[ 1+\ln \left( 2\pi Q/\beta \right) %
\right] \right) }{4\pi l^{2}}-\frac{2\beta ^{2}r_{+}\ln \left( r_{+}\right) 
}{\pi }+O\left( r_{+}^{3}\right) ,  \label{temexp}
\end{equation}%
where $Q_{m}=\left( 1+c_{0}^{2}c_{2}m^{2}\right) /\left( 16\pi \beta \right) 
$ is the `\textit{marginal }charge' and we use $\Lambda =-3/l^{2}$. As one
can see, the behavior of the temperature for small $r_{+}$ depends on the
charge of the black hole. For $Q>Q_{m}$, black hole has zero temperature for
a specific value of $r_{+}$ and for values greater than it temperature is
positive. It means that these black holes are Reissner-Nordstrom (RN) type.
For $Q<Q_{m}$, the black hole does not exist in low temperature region and
temperature diverges as $r_{+}$ goes to zero (similar to Schwarzschild
solution). So, these black holes are Schwarzschild (S) type. Note that for $%
Q=Q_{m}$, the black hole has a finite positive temperature as $r_{+}$ tends
to zero. Therefore, the latter is neither S-type, nor RN-type and is
marginal. This is the effect of massive term $c_{0}c_{1}m^{2}/4\pi $ in (\ref%
{temexp}). For massless case ($m=0$), $T\rightarrow 0$ as $r_{+}\rightarrow
0 $ for $Q=Q_{m}$ and the black hole is RN-type \cite{1711.01151}. The large 
$r_{+}$ limit of the temperature is $3r_{+}/4\pi l^{2}$ which is clearly
independent of the charge and increases linearly as $r_{+}$ increases (see
Fig. \ref{figure13}).

Now, we intend to obtain the critical point (where continuous phase
transition occurs). In order to characterize the critical point, we have%
\begin{equation}
\frac{\partial T}{\partial S}\Big|_{Q_{c}}=0,\quad \quad \quad \frac{%
\partial ^{2}T}{\partial S^{2}}\Big|_{Q_{c}}=0,  \label{cpoint}
\end{equation}%
where $Q_{c}$ is the critical charge. Using (\ref{cpoint}), one could
determine $Q_{c}$ and $r_{+c}$ and specify the critical point ($%
Q_{c},T_{c},r_{+c}$). In Fig. \ref{figure14}, we plot the behavior of $Q_{c}$%
, $T_{c}$ and $r_{+c}$ with respect to $\beta $ for a set of model
parameters by solving the equations of (\ref{cpoint}) numerically. The
vertical dashed lines in Fig. \ref{figure14} show the range in which the
critical behavior takes place for S-type black holes. For $\beta $ values
larger than one specified by the rightmost vertical dashed line, the
critical point occurs for RN-type black holes.

Now, we turn to study the behavior of Gibbs free energy for our black holes
to find out the phase structure of the system. It determines the globally
stable state at equilibrium process. The lowest Gibbs free energy specifies
the global stable state. One could obtain the Gibbs free energy per unit
volume $V_{2}$ as \cite{RPT1}%
\begin{equation}
G=M-TS,
\end{equation}%
We focus on RN-type ($3+1$)-dimensional logarithmic charged massive black
holes. The Gibbs free energy as a function of temperature for different
values of charge is illustrated in Fig. \ref{figure15}. In addition, the
phase diagram is depicted in the $Q$--$T$ diagram in Figs. \ref{figure16}.
According to the value of charge $Q$, we have different phase transitions.
For $Q=Q_{c}$, the system is at critical point. This point is highlighted by
a black spot in Fig. \ref{figure16}. For charge values greater than $Q_{c}$,
the system is locally stable ($C_{Q}>0$) and Gibbs free energy is single
valued. Therefore, no phase transition happens. For charge values less than
critical value $Q_{c}$, there are thermally instable phases in the system ($%
C_{Q}<0$). So, the system experiences different phase transitions depending
on the value of charge $Q$. A first order phase transition between small and
large black holes occurs for $Q_{z}<Q<Q_{c}$. This is accompanied by a
discontinuity in the slope of Gibbs free energy at transition point. This
phase transition is marked by the gold curve in Fig. \ref{figure16}. For $Q$
in the range of $Q_{t}<Q<Q_{z}$, an interesting phenomenon happens. In this
range, in addition to a first order phase transition which sepatates small
and large black holes, a zeroth order phase transition between small and
intermediate black holes occurs. This can be seen as a finite jump in Gibbs
free energy in Fig. \ref{figure15}. The latter initiates from $Q=Q_{z}$ and
terminates at $Q=Q_{t}$. $z$ and $t$ points are specified in Fig. \ref%
{figure16}. In this figure, this zeroth order phase transition is identified
by purple curve. This intermediate (large)/small/large transition shows a
reentrant phase transition \cite{RPT1,RPT2}. This is the first observation
of reentrant phase transition for black hole solutions in massive gravity
without extending thermodynamics phase space. In \cite{RPT3}, the occurance
of reentrant phase transition in extended thermodynamics phase space has
been reported for black holes in massive gravity. As Fig. \ref{figure15}
exhibits, no phase transition happens for $Q<Q_{t}$ and only stable large
black holes exists. Also, no black hole region in Fig. \ref{figure16}
corresponds to no black hole solution exists.

\section{Summary and CLOSING REMARKS}

The importance of massive gravity model is at least twofold. On the one
hand, because of the recent developments in gravitational waves observatory
indicating the speed of gravitational wave deviates from the speed of light
which implies the existence of massive graviton, we need an extension of
general relativity with a massive graviton as intermediate particle similar
to $W$ and $Z$ in electroweak interaction. On the other hand, from
holographic viewpoint, giving mass to gravity could model momentum
dissipation in dual system on boundary. In this work, we have studied black
hole solutions in massive gravity model in the presence of LNE Lagrangian.
As well as BI model, in the logarithmic one, there is no divergency in the
electric field of a point particle at the origin.

We first constructed the field equations by varying the action. Then, we
obtained the exact solution of the metric function and studied the behavior
of the obtained solutions. We observed that depending on the model
parameters, it is possible to have two horizons, an extreme horizon, or
naked singularity. Moreover, one of the effects of massive gravity is the
existence of multi horizons. This phenomenon is due to anti-evaporation
which is a quantum effect. Next, we calculated the conserved and
thermodynamic quantities. Calculation of these quantities is necessary for
checking the satisfaction of the first law of black hole thermodynamics. It
was observed that in the presence of nonlinear electrodynamics and within
the context of massive gravity model, the first law of thermodynamic is
still valid.

We studied the thermal stability in both canonical and grand canonical
ensembles. We observed that considering different values for the model
parameters leads to a variety of phase structures for our black holes. In
canonical ensemble, the heat capacity can experience both positive and
negative values. The position of its root(s) and divergency(ies) affected by
metric parameters. In this ensemble, our black holes encounter two types of
phase transitions; the second order phase transition in the position of
divergences of heat capacity with positive temperature, and transitions
between un-physical ($T<0$) and physical ($T>0$) states of solutions.
Increasing the values of electric charge ($Q$) or massive parameter ($m$)
causes the transition between un-physical and physical black holes in larger
sizes (Fig. \ref{fig4}). It means that there is a lower bound on black hole
radius increased as $Q$ and $m$ are enhanced. Also, by decreasing the
nonlinearity parameter ($\beta $), second order phase transition happens for
smaller black holes (Fig. \ref{fig5}). In the grand canonical ensemble, the
determinant of the Hessian matrix $\mathbf{H}$ had a root with positive
temperature, which its value depends on the metric parameters. We found out
that some horizon radii are not allowed for our solution since $\mathbf{H}$
is negative for them. For instance, when other parameters are fixed, the
stability is observed for larger black hole when massive parameter $m$ (or
dimension $n$) increases (Fig. \ref{fig9}). Also, for smaller values of
electric charge, the stability is guaranteed for larger black holes (Fig. %
\ref{fig10}).

Finally, we investigated the critical behavior of ($3+1$)-dimensional black
holes with spherical event horizon. We showed that there is a marginal
charge $Q_{m}$ that for charge values greater than it black holes are
Reissner-Nordstrom-type, whereas for charge values less than it black holes
are Schwarzschild-type. For massless gravity, the black hole solutions with $%
Q=Q_{m}$ are RN-type. We observed that in the context of massive gravity the
latter are neither S-type nor RN-type and have a marginal behavior due to
the effects of massive term. Next, we focused on RN-type black holes and
study the behavior of Gibbs free energy to disclose the phase structure of
solutions. We found out that there are different phase transitions according
to the value of charge $Q$. There are some ranges for which there is no
phase transition. However, for another range we just have a first order
phase transition between small and large black holes accompined by a
discontinuity in the slope of Gibbs free energy at transition point.
Interestingly enough, there is a range of charge values for which black
holes experience a zeroth order phase transition where a finite jump in
Gibbs free energy value occurs. This causes a reentrant phase transition
i.e. a large/small/large phase transition. This is the first report of
reentrant phase transition for black holes in massive gravity without
needing to extend thermodynamics phase space.

The study of this work can be extended from differnt aspects. For instance,
in this context, one could explore the behavior of holographic conductivity.
Furthermore, the effects of massive gravity with logarithmic electrodynamics
on holographic superconductors could be investigated. Here, we examined the
critical behavior of ($3+1$)-dimensional black holes numerically. This may
be studied analytically. Also, one could explore this behavior for higher
dimensional black holes. Moreover, the effects of more model parameters on
phase behavior of such systems could be taken under study. Some of these
pointed out issues are now under investigation.

\begin{acknowledgments}
AS and SMZ thank the Research Council of Shiraz University. MKZ would like
to thank Shahid Chamran University of Ahvaz for supporting this work. We
gratefully acknowledge useful discussions with A. Dehyadegari.
\end{acknowledgments}

\end{document}